\documentclass[nofootinbib,amsmath,amssymb,showpacs,showkeys,aps,reprint]{revtex4-1}
\usepackage[utf8,latin1]{inputenc}
\usepackage{graphicx}
\usepackage{dcolumn}
\usepackage[dvipsnames]{xcolor}
\usepackage[T1]{fontenc}

\usepackage{mathrsfs}  
\usepackage{cases}
\usepackage{bm}
\usepackage{academicons}
\usepackage{braket}
\usepackage{booktabs}
\usepackage{comment}
\usepackage{caption}
\usepackage{mathtools, nccmath}
\usepackage{fancyhdr}
\usepackage{tikz,xcolor}
\usepackage{tensor}
\usepackage[normalem]{ulem}
\usepackage{lipsum}
\usepackage{soul}
\usepackage{cancel}
\usepackage{accents}
\usepackage{stackengine,scalerel}
\usepackage{hyperref}
\usepackage{tabularx}
\hypersetup{colorlinks, linkcolor={red},citecolor={blue},urlcolor={blue}}

\def\nn{\nonumber} 
\usepackage{calrsfs}
    \DeclareMathAlphabet{\pazocal}{OMS}{zplm}{m}{n}
    
    \newcommand{\defeq}{\vcentcolon=}
    \newcommand{\del}{\partial}
    
    \newcommand{\dd}{{\rm d}}
    
    \newcommand{\R}{\mathbb{R}}

    \newcommand{\ii}{{\rm i}}
    
    \newcommand{\bt}{\hat{B}}
     
    \newcommand{\lc}[1]{\accentset{\circ}{#1}}
    \newcommand{\nm}[1]{\accentset{\diamond}{#1}}
    \newcommand{\qm}[1]{``#1''}
    \definecolor{mycolor}{rgb}{0.0, 0.5, 0.13}

\definecolor{lime}{HTML}{A6CE39}
\DeclareRobustCommand{\orcidicon}{
	\begin{tikzpicture}
	\draw[lime, fill=lime] (0,0) 
	circle [radius=0.16] 
	node[white] {{\fontfamily{qag}\selectfont \tiny ID}};
	\draw[white, fill=white] (-0.0625,0.095) 
	circle [radius=0.007];
	\end{tikzpicture}
	\hspace{-2mm}
}

\fancypagestyle{plain}{%
  \fancyhf{}
  \fancyfoot[C]{\iffloatpage{}{\thepage}}
  }
\foreach \x in {A, ..., Z}{%
	\expandafter\xdef\csname orcid\x\endcsname{\noexpand\href{https://orcid.org/\csname orcidauthor\x\endcsname}{\noexpand\orcidicon}}
}

\begin{document}

\title[Minisuperspace  Cosmology in Extended Geometric Trinity of Gravity]{Minisuperspace  Cosmology in Extended Geometric Trinity of Gravity}

\author{Emmanuele Battista\orcidA{}$^{1}$\vspace{0.5cm}}\email{ebattista@lnf.infn.it} 
\author{Salvatore Capozziello\orcidC{}$^{2,3,4}$}\email{capozziello@na.infn.it}
\author{Stefano Pastore\orcidB{}$^{2,3}$\vspace{0.5cm}} \email{stefano.pastore@na.infn.it}

\affiliation{$^1$ Istituto Nazionale di Fisica Nucleare, Laboratori Nazionali di Frascati, I-00044 Frascati, Italy, \\
$^2$ Dipartimento di Fisica ``Ettore Pancini'', Complesso Universitario 
di Monte S. Angelo, Universit\`a degli Studi di Napoli ``Federico II'', Via Cinthia Edificio 9, I-80126 Napoli, Italy,\\
$^3$ Istituto Nazionale di Fisica Nucleare, Sezione di Napoli, Complesso Universitario 
di Monte S. Angelo, Via Cinthia Edificio 9, I-80126 Napoli, Italy,\\
$^4$ Scuola Superiore Meridionale, Via Mezzocannone 4, I-80134 Napoli, Italy.}

\date{\today} 

\begin{abstract}

We investigate Extended Geometric Trinity of Gravity  at both classical and quantum cosmological levels using the minisuperspace  approach. Adopting  Noether symmetries  to select viable models,  we examine   metric-affine theories of gravity, in particular the extensions of General Relativity, Teleparallel Equivalent General Relativity and Symmetric Teleparallel Equivalent General Relativity,  and show that the equivalence among these different formulations can be restored by including in the Lagrangian the divergence terms that relate their respective geometric invariants to the Ricci scalar. Exact cosmological solutions are derived and compared in the different models.

\end{abstract}

\maketitle

{
\hypersetup{linkcolor=black}
\tableofcontents
}

\section{Introduction}

General Relativity (GR)  is a metric theory where dynamics is described by the Riemann tensor. The physical basis of GR is the Equivalence Principle in its various formulations and, thanks to it, the causal structure (related to the metric) and the geodesic structure (related to the free fall) coincide. The immediate consequence of this physical principle is the fact that the affine connection is unequivocally defined, that is  the Levi-Civita  connection is derived from a non-linear combination of metric and its derivatives.  However,  the question of why all the degrees of freedom of spacetime should be related only to the metric remains unanswered up to now. On the other hand, GR is a robust theory when compared with experiments and astronomical observations. For  this reason,  equivalent formulations of gravity, where the metric emerges from some more fundamental geometric  structure,  represents a promising path towards a deeper understanding of  gravitational interaction \cite{Krasnov_2020, Meluccio.pregeom.2024, Meluccio.pregeom.hamilt.2025}.
    
    One of the earliest attempts to explore this idea was carried out by \`Elie Cartan in 1922 \cite{Cartan.1922}, by introducing torsion into the theory via the tetrad formalism. A related approach was pursued by Einstein himself six years later, motivated by the fact that the tetrad field contains sixteen components, compared to the ten independent components of the metric tensor. Einstein  then conjectured that the six additional degrees of freedom might account for the electromagnetic field. However, this proposal was unsuccessful, as the extra degrees of freedom associated with the tetrad are pure gauge and  can be eliminated by the local six-parameter Lorentz transformations. This approach to gravity was  revisited forty years later, independently by Kibble \cite{Kibble.1961} and Sciama \cite{Sciama.1964}, in an effort to formulate gravity as a gauge theory of the Poincar\'e group.
    
    At the same time, M\o ller showed that a tetrad-based description of gravity  allowed for a more natural definition of gravitational energy than in the standard GR \cite{Moller.1961, Moller.1961.b}. Following the same lines,  a new formulation of gravity as a gauge theory of translations emerged \cite{Pellegrini.1963, Hayashi.Nakano.1967, Hayashi.1979}. This approach   is  now referred to as \textit{Teleparallel Gravity} \cite{aldrovandi2012teleparallel, LeDelliou.tg.2019, Bahamonde_2023}. A few years later, yet another scenario appeared, called  \textit{Symmetric Teleparallel Gravity}, where both curvature and torsion vanish, with the non-metricity tensor being the main geometric object  encoding the gravitational interaction \cite{Nester.1998, Adak.stg.2006, Jimenez.2017}. Teleparallel Gravity and Symmetric Teleparallel Gravity have  been shown to be dynamically equivalent to GR, and  hence these three formulations  are now collectively  regarded  as the \textit{Geometric Trinity of Gravity} \cite{Jimenez.trinity, capozziello-defalco-ferrara, Mancini:2025asp}.

    These theories admit   extensions. Since no fundamental principle prescribes the exact form of the Lagrangian beyond   symmetry invariance, one can consider modified versions of each model. This leads to the framework of the extended General Relativity \cite{Capozziello:2002rd,   Nojiri.2006, Nojiri.2010, DeFelice:2010aj, Nojiri.2017}, and similar extensions can be worked out for the other two theories of the geometric trinity \cite{Harko.2011, Cai.2015,  Nojiri.2017, DAmbrosio.heis.bhfq.2022, Heisenberg_fQ, Capriolo.2024}. The ensuing  gravity paradigms are no longer equivalent and the correspondence can be restored by taking into account appropriate  pure divergence terms in the Lagrangian. The resulting class of theories is  referred to as the \textit{ Extended Geometric Trinity of Gravity} \cite{Capozziello-ext-trinity-2025, Hess.ghy}. Clearly, also if dynamics are equivalent, the foundation and possible experimental paradigms to test them can be different  \cite{Mancini:2025asp}.

    Both  reformulations and extensions of GR  can be fundamentally relevant to the quest for  Quantum  Gravity. It is well known that GR, in its standard form, faces serious challenges under conventional quantization schemes. One of the main obstacles lies on the fact that Quantum Field Theory is typically formulated on a fixed Minkowski background, whereas GR treats the metric itself as a dynamical entity. Consequently, GR turns out to be a non-renormalizable quantum  theory \cite{maggiore.qft}, with the renormalization procedure inevitably introducing higher-order curvature invariants into the Lagrangian \cite{Birrell:1982ix, capozziello.beyond.2011}. This fact motivates the investigation of extended or modified gravitational frameworks that might offer a way towards  the quantization of gravitational field. One of such a possibility is to take into account  approaches where the metric is no longer treated as a fundamental variable, but rather as an emergent quantity stemming from more fundamental geometric structures \cite{Meluccio.pregeom.2024, Meluccio.pregeom.hamilt.2025}. 

    In this paper, we aim to explore  cosmological aspects  of Extended Geometric Trinity of Gravity, considering  the so-called \textit{minisuperspace} approach \cite{Isichei2022, Capozziello:1999xr, Capozziello:2024clc, Capozziello:2022vyd}. This framework leads  naturally to the formalism of quantum cosmology \cite{wiltshire.2003.QC}. The goal of this paper is to demonstrate that the  equivalence among the different gravity formulations within the Extended Geometric Trinity can be recovered   also for minisuperspace quantum cosmology.

    The plan of the  paper is as follows. We  review the main aspects of  metric-affine theories  of gravity in Sec. \ref{section.overview.metric.affine}.  Sec. \ref{section.noether} is devoted to a   brief discussion of the canonical quantization of gravity and the Noether Symmetry Approach. The method is then applied, in Sec. \ref{section.fR}, to  $f(R)$ gravity. A similar analysis is carried out in Sec. \ref{section.fQ.fQB} for non-metric gravity. There, in \ref{subsec.fQ}, we first show that the $f(Q)$ cosmological solutions are not equivalent to the $f(R)$ ones. In Sec. \ref{subsec.fQB}, we   introduce a pure divergence term into the $f(Q)$ Lagrangian and analyze the specific cases where the equivalence with $f(R)$ gravity is restored. In Sec. \ref{section.fT.fTB}, we examine     $f(T)$ and $f(T,\bt)$ teleparallel gravity   showing the equivalence with respect to the   metric and non-metric  cases. Discussion and  conclusions are drawn in Sec. \ref{section.conclusions}. In Table I, a comprehensive summary of models and solutions is reported.

\section{The geometric trinity of gravity} \label{section.overview.metric.affine}

     Einstein theory describes the dynamics of metric tensor $g_{\mu\nu}$ on a smooth manifold $\pazocal{M}$. The geometric structure of spacetime is thus encoded in the pair $(\pazocal{M},g)$, where $g$ determines both the causal structure and the geodesic structure. In this framework, the affine connection is  the Levi-Civita one, whose components read as 
    \begin{align} \label{lc.connection}
        \tensor{\lc{\Gamma}}{^\lambda_\mu_\nu}=\frac{1}{2}g^{\lambda\rho}(\del_\mu g_{\rho\nu}+\del_\nu g_{\rho\mu}-\del_\rho g_{\mu\nu})\,.
    \end{align}
    Since $\tensor{\lc{\Gamma}}{^\lambda_\mu_\nu}$ is completely determined by the metric tensor and its derivatives, GR is classified as a \textit{metric theory of gravity}.

    In the broader context of the \textit{Metric-affine Theories of Gravity} (MAG),
    the triple $(\pazocal{M}, g, \Gamma)$ determines geometry and dynamics. In this setting,  the affine connection is no longer specified solely in terms of  metric, but it takes the general form 
    \begin{align} \label{general.connection}
            \Gamma^\lambda_{\mu\nu} = \lc\Gamma^{\lambda}_{\mu\nu} + \tensor{K}{^\lambda_\mu_\nu} + \tensor{L}{^\lambda_\mu_\nu}\,,
    \end{align}
    where  the \textit{contortion} and \textit{disformation} tensors are defined, respectively, as
    \begin{subequations}
        \begin{align}
            &\tensor{K}{^\lambda_\mu_\nu} \defeq \frac{1}{2}(\tensor{T}{_\nu^\lambda_\mu}+\tensor{T}{_\mu^\lambda_\nu}-\tensor{T}{^\lambda_\mu_\nu})\,,\\
            &\tensor{L}{^\lambda_\mu_\nu}\defeq\frac{1}{2}(\tensor{Q}{^\lambda_\mu_\nu} - \tensor{Q}{_\mu^\lambda_\nu}-\tensor{Q}{_\nu^\lambda_\mu})\,,\label{disformation.tensor}
        \end{align}
    \end{subequations}
    with 
     \begin{subequations}
        \begin{align}
            &\tensor{T}{^\lambda_\mu_\nu}\defeq \tensor{\Gamma}{^\lambda_\nu_\mu} - \tensor{\Gamma}{^\lambda_\mu_\nu}\,, \label{torsion.comp} \\
            &Q_{\lambda\mu\nu}\defeq\nabla_\lambda g_{\mu\nu}\,, \label{non.metricity} 
        \end{align}
    \end{subequations}
    being the \textit{torsion} and \textit{non-metricity} tensors, respectively. The Riemann tensor
    \begin{align}
        \tensor{R}{^\alpha_\beta_\mu_\nu}\defeq &\, \del_\mu\tensor{\Gamma}{^\alpha_\beta_\nu} - \del_\nu\tensor{\Gamma}{^\alpha_\beta_\mu} + \tensor{\Gamma}{^\alpha_\lambda_\mu}\tensor{\Gamma}{^\lambda_\beta_\nu} - \tensor{\Gamma}{^\alpha_\lambda_\nu}\tensor{\Gamma}{^\lambda_\beta_\mu}
    \end{align}
    can be decomposed as
    \begin{align}
        \tensor{R}{^\alpha_\beta_\mu_\nu} = & \tensor{\lc{R}}{^\alpha_\beta_\mu_\nu} + \lc{\nabla}_\mu\tensor{M}{^\alpha_\beta_\nu} - \lc{\nabla}_\nu\tensor{M}{^\alpha_\beta_\mu} \nn \\
        &+ \tensor{M}{^\alpha_\lambda_\mu}\tensor{M}{^\lambda_\beta_\nu} - \tensor{M}{^\alpha_\lambda_\nu}\tensor{M}{^\lambda_\beta_\mu}\,,
    \end{align}
    where $\tensor{\lc{R}}{^\alpha_\beta_\mu_\nu}$ and $\lc{\nabla}$ are the Riemann tensor and the covariant derivative relative to the Levi-Civita connection \eqref{lc.connection}, respectively, and $\tensor{M}{^\alpha_\beta_\nu}:=\tensor{K}{^\alpha_\beta_\nu}+\tensor{L}{^\alpha_\beta_\nu}$.

    The Geometric Trinity of Gravity  is  a subclass of MAG. We   provide a classification of these models in Sec. \ref{subsec.mag.classification}, in view of the analysis presented  in Sec. \ref{subsec.geometric.trinity}.

    \subsection{Classification of metric-affine theories}\label{subsec.mag.classification}

\begin{figure}[bht!]
            \centering
            \includegraphics[width=8.3 cm]{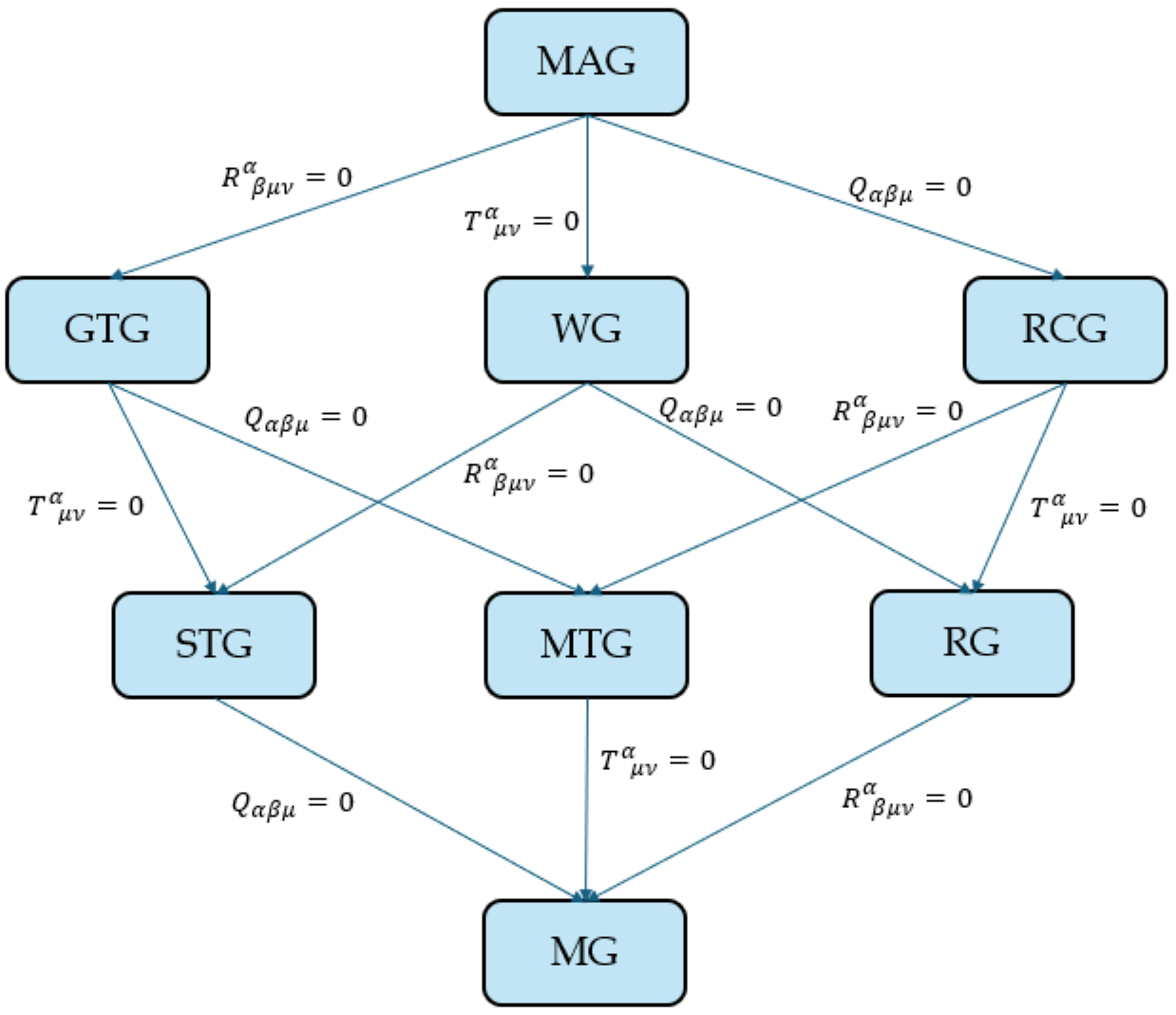}
            \caption{Classification of MAGs. RG, MTG, and STG define the three geometrical realizations of the \emph{Geometric Trinity of Gravity}.}
            \label{fig:MAT}
        \end{figure}

        The fundamental geometric objects of  MAG  are the  curvature, torsion and non-metricity tensors. It is however convenient to study  cases where one or more of these tensors vanish (see Fig.  \ref{fig:MAT}): 
        \begin{enumerate}
            \item $Q_{\lambda\mu\nu}=0$: Riemann-Cartan geometry (RCG). This is also known as the  Einstein-Cartan theory, in which the torsion tensor is related to the  spin  \cite{Hehl.Einstein.Cartan}. It was  introduced by Cartan in 1922, and  later explored by Einstein, following his unsuccessful attempts to unify GR with electromagnetism \cite{history.uft}. 
            \item $\tensor{T}{^\lambda_\mu_\nu}=0$: Weyl geometry (WG). This theory was  formulated by Weyl to unify gravity with electromagnetism within a fully geometric framework \cite{Wheeler-Weyl.geom}. It emerges as the gauge theory of the Weyl group \cite{Condeescu.weyl}. 
            \item $\tensor{R}{^\alpha_\beta_\mu_\nu}=0$: General teleparallel geometries (GTG). These theories are formulated in terms of  torsion and non-metricity tensors, and  constitute an alternative to GR. Their most remarkable feature is the notion of ``absolute parallelism'', or \textit{teleparallelism}: given two vectors belonging to two different tangent spaces, it is possible to parallel transport one into the tangent space of the other, with the outcome not depending on the chosen path. As a result, the notion of parallelism at  distance can be established. GTG are further divided into:
            \begin{itemize}
                \item $Q_{\lambda\mu\nu}=0$: metric teleparallel geometries (MTG);
                \item $\tensor{T}{^\lambda_\mu_\nu}=0$: symmetric teleparallel geometries (STG).
            \end{itemize}
            \item $Q_{\lambda\mu\nu}=0\,,\,\tensor{T}{^\lambda_\mu_\nu}=0$: Riemannian geometry (RG) \cite{lee.riemann.2019}. This is the arena of GR, which is  built solely upon the curvature tensor \cite{Wald:1984rg, misner2017gravitation}.
            \item The Minkowski geometry (MG) is obtained by setting all the three tensors $\tensor{R}{^\alpha_\beta_\mu_\nu}$, $\tensor{T}{^\lambda_\mu_\nu}$ and $Q_{\lambda\mu\nu}$ to zero, and represents the geometric framework of Special Relativity.   
        \end{enumerate}

    \subsection{Equivalence of metric-affine theories} \label{subsec.geometric.trinity}

        Among the various metric-affine theories of gravity, GR and those based on teleparallel geometries are of particular interest. In this paper, we will  focus on the  Teleparallel Equivalent of General Relativity (TEGR) and the Symmetric Teleparallel Equivalent of General Relativity (STEGR), which arise from MTGs and STGs, respectively.

        One of the fundamental features of GR is its \textit{universality}, which means that all forms of matter and energy are affected by gravity in the same way. This property is due to the Equivalence Principle. In this framework, the gravitational effects are encoded in the Riemann curvature tensor derived from the Levi-Civita connection \eqref{lc.connection}. Test particles in free fall follow geodesic trajectories determined by the spacetime geometry \cite{Ehlers:2012wgb}.
        
        In theories based on teleparallelism, this framework changes. In TEGR, gravity is described by the torsion tensor \eqref{torsion.comp}, and the gravitational interaction arises as a (gauge) force. Here, the equations of motion for freely falling objects are described by the force equations \cite{Bahamonde_2023, Iosifidis.2024eom}
        \begin{align}\label{tegr.geodesic}
            \frac{\dd^2x^\lambda}{\dd s^2} + \hat\Gamma^\lambda_{\mu\nu}\frac{\dd x^\mu}{\dd s} \frac{\dd x^\nu}{\dd s} = \tensor{\hat{K}}{^\lambda_\mu_\nu} \frac{\dd x^\mu}{\dd s} \frac{\dd x^\nu}{\dd s}\,,
        \end{align}
        where it is clear that the contortion tensor enters the equations of motion as a force field. STEGR is somewhat similar to TEGR: the geometric object that characterizes gravity is the non-metricity tensor \eqref{non.metricity}, and the equations of motion take the form \cite{Lemos.2024}
        \begin{align}\label{stegr.geodesic}
            \frac{\dd^2x^\lambda}{\dd s^2} + \nm\Gamma^\lambda_{\mu\nu}\frac{\dd x^\mu}{\dd s} \frac{\dd x^\nu}{\dd s} = \tensor{\nm{L}}{^\lambda_\mu_\nu} \frac{\dd x^\mu}{\dd s} \frac{\dd x^\nu}{\dd s}\,.
        \end{align}
        
        Despite the conceptual differences between GR and teleparallel theories, it has been shown that they provide  equivalent dynamics and the Einstein-Hilbert action can be recovered considering boundary terms for the torsion scalar $T$ and the non-metricity scalar $Q$ \cite{Jimenez.trinity, capozziello-defalco-ferrara}. Taking into account suitable boundary terms    the generalization to the Extended Geometric Trinity of Gravity directly emerges \cite{Capozziello-ext-trinity-2025}. The minisuperspace cosmology of such  extensions will be studied in Sec. \ref{section.fQ.fQB} and \ref{section.fT.fTB}.

\section{Quantum cosmology and Noether symmetries} \label{section.noether}

    Let us discuss now the connection between the quantum cosmology formalism in the minisuperspace framework and Noether's theorem, which plays a crucial role in the analysis of the following sections.
    
    A possible method for  quantizing  gravity is given by the so-called {\it canonical approach}.  It relies on the Arnowitt-Deser-Misner (ADM) formalism \cite{ADM}, which provides a Hamiltonian formulation of GR \cite{dirac1964, regge1976constrained, DeWitt-1967}. The quantization procedure consists of  replacing the momenta by functional derivatives with respect to the conjugate quantities in the  Hamiltonian and momentum constraint equations of classical GR. The main result of the ADM approach is the   Wheeler--De Witt equation, whose solution,  the  \textit{Wave Function of the Universe}, is related to   the probability that  observable universes emerge \cite{Hawking.Halliwell.1985}.

    The Wheeler--DeWitt equation is a second-order functional differential equation defined on an infinite-dimensional space known  as \textit{superspace}. Up to now, no general procedure exists for solving it analytically. A possible way to address this issue is provided by the so-called \textit{minisuperspace approach} to quantum cosmology, where most of the degrees of freedom in the superspace are fixed by imposing that the metric satisfies a given symmetry \cite{hawking.luttrell.1984}. As a result, the phase space becomes  finite dimensional, and the Wheeler--DeWitt equation reduces to  a second-order differential equation, which can be, in principle,    analytically solved \cite{Isichei2022}. In this paper, we will apply the  minisuperspace approach to cosmological models derived from MAG theories. 
    
    While the minisuperspace approach has shown to have predictive power, the whole theory still lacks a rigorous approximation procedure from superspace. In particular, it could be argued that it violates the Heisenberg Uncertainty Principle due to the freezing of the majority of its degrees of freedom along with its conjugate momenta. On the other hand, spatial homogeneity and isotropy are important approximations in classical cosmology with solid observational basis, so it is  reasonable to assume that a truncation procedure from the full quantum cosmology theory would result in such models. Moreover, it has been shown that the minisuperspace formalism is compatible with  the Quantum Field Theory on curved spaces \cite{Birrell:1982ix, Hawking.Halliwell.1985, vilenkin1989}, further motivating cosmologists to explore these models. 

    A particularly intriguing aspect of  minisuperspace  approach arises from the existence of Noether symmetries. They behave as a  rule to select  viable dynamical systems  related to observable universes. See Refs. \cite{Capozziello:1999xr, capozziello.odintsov.hamiltonian, bajardi.book.noether} for a discussion.

    \subsection{The Noether Symmetry Approach}\label{subsec.nsa}
    
        As pointed out before, the phase space related to  minisuperspaces  presents a finite number of degrees of freedom. Therefore, let us  recall some concepts of Noether theorem for a point-like Lagrangian, which, in the next sections, will depend on the scale factor $a$ and the geometric scalar  characterizing the specific theory of gravity under consideration. 
        
        We assume  a non-degenerate  Lagrangian function 
        \begin{align}
            \pazocal{L}(q,\dot q) = K(q,\dot q) - V(q)\,,
        \end{align}
        for which the Hessian determinant satisfies the condition
        \begin{align}
        \det H_{ij}\defeq&\det{\frac{\del^2\pazocal{L}}{\del\dot q^i\del\dot q^j}}\neq0\,.\label{non.singularity.condition}
        \end{align}
        Here, $q=\{q^i\}_{i=1}^n$ are the generalized coordinates describing the $n$ degrees of freedom of the system, the dot stands for the derivative with respect to time. Then $\dot q^i$ are the generalized velocities. $K$ and $V$ are the kinetic and potential energy, respectively. 
        
        We further suppose the Lagrangian to be time independent, i.e.,
        \begin{align}
            \frac{\del\pazocal{L}}{\del\tau}=&0\,, \label{time.independent.lagrangian}
        \end{align}
        which implies that  the associated energy function 
        \begin{align}\label{energy.from.lagrangian}
            E_{\pazocal{L}}\defeq\frac{\del\pazocal{L}}{\del\dot q^i}\dot q^i - \pazocal{L} = K+V\,,
        \end{align}
        is a constant of motion \cite{landau1976mechanics}. 
        
        Since  cosmological models have finite degrees of freedom, we restrict our attention to non-degenerate pointlike transformations of the form
        \begin{align}\label{point.transformation}
            Q^i=Q^i(q)\,,
        \end{align}
        which induce the following transformation on the generalized velocities:
        \begin{align}
            \dot Q^i = \frac{\del Q^i}{\del q^j}\dot q^j\,.
        \end{align}
    
        Let us now consider a one-parameter family of transformations depending on a small parameter $\varepsilon$:
        \begin{align}
            Q^i=Q^i(q, \varepsilon).
        \end{align}
        For infinitesimal $\varepsilon$, we can define the \textit{generators of the transformation} as \cite{Capozziello-DeFelice}
        \begin{align}\label{generators}
            X\defeq\alpha^i(q)\frac{\del}{\del q^i} + \dot\alpha^i(q)\frac{\del}{\del\dot q^i}\,,
        \end{align}
        which defines  a vector field on the tangent bundle $T\pazocal{Q}\defeq\{q,\dot{q}\}$,  with $\pazocal{Q}\defeq\{q\}$  the configuration space,  whose integral curves correspond to the orbits of  transformation $\bigl(Q^i=Q^i(q_0, \varepsilon), \dot Q^i=\dot Q^i(q_0, \varepsilon)\bigr)$, where $q_0=\{q^i_0\}$ is fixed. 
    
        A function $F(q, \dot q)$ is said to be invariant under the transformation generated by $X$ if its Lie derivative
        \begin{align}
            L_XF \equiv XF
        \end{align}
        vanishes. If such function coincides with the Lagrangian, then we can write 
        \begin{align}\label{symmetry.definition}
            L_X\pazocal{L}=0\,,
        \end{align}
        and the transformation is called a \textit{symmetry} of the system, generated by the vector field $X$.

        \textit{Noether's theorem}  states that if condition  \eqref{symmetry.definition}, together with the Euler--Lagrange equations, hold, then the quantity 
        \begin{align}\label{noether.charge}
            j_0\defeq\alpha^i\frac{\del \pazocal{L}}{\del\dot q^i}
        \end{align}
        is a constant of motion, referred to as the conserved \textit{Noether charge}. 
        
        Under the transformation \eqref{point.transformation}, the vector field \eqref{generators} changes according to
        \begin{align}
           X \rightarrow \tilde{X}=(i_X\dd Q^k)\frac{\del}{\del Q^k} + \biggl(\frac{\dd}{\dd\tau}(i_X\dd Q^k)\biggr)\frac{\del}{\del\dot Q^k}\,,
        \end{align}
        where $i_X$ is the inner product of forms \cite{Nakahara.2003,Capozziello-DeFelice}. If $X$ is the generator of a symmetry, one can choose a transformation such that
        \begin{subequations}\label{reduction.condition}
            \begin{align}
                &i_X\dd Q^1=1\,, \\
                &i_X\dd Q^k=0\,\,\text{for}\,\,k\neq1\,,
            \end{align}            
        \end{subequations}
        which yields 
        \begin{align}
            \tilde{X}=\frac{\del}{\del Q^1}\,,
        \end{align}
        and hence   the symmetry condition takes the simple form 
        \begin{align}\label{cyclic.condition}
            \frac{\del\pazocal{L}}{\del Q^1}=0\,.
        \end{align}
        If such a relation is satisfied, $Q^1$ is a \textit{cyclic} variable, and the system is said to be \textit{reduced}; in this case, it  follows from the Lagrange equations that
        \begin{align}
          \pi_1=  \frac{\del\pazocal{L}}{\del\dot Q^1} = j_0 = \text{constant}\,,
          \label{pi-1-const}
        \end{align}
        in agreement with Noether's theorem.
    
        Noether's theorem is of remarkable importance in quantum cosmology. In this framework, the quantization procedure establishes that  the classical condition \eqref{pi-1-const} translates into the supplementary equation
        \begin{align}
            \pi_1\Psi(Q^1,...,Q^n)=j_0\Psi(Q^1,...,Q^n)\,,
        \end{align}
        which allows to factorize the the wave function $\Psi$ into
        \begin{align}
            \Psi(Q^1,...,Q^n)=e^{\ii j_0Q^1}\phi(Q^2,...,Q^n)\,.
        \end{align}
        The Noether Symmetry Approach thus provides a realization of the Hartle criterion \cite{Hartle.crit}, which states  that  correlations among the variables of the Wave Function of the Universe indicate the possibility to obtain classically observable universes. In particular, this happens as soon as oscillatory behaviors are present.  Conversely, in regions where $\Psi$ exhibits  exponential behaviors,  correlations are  not present and classically observable  spacetimes cannot emerge \cite{Capozziello:1999xr}. 

        This interpretation is further supported in the semiclassical limit, where the wave function can be approximated as \cite{maggiore.qft}
        \begin{align}\label{wfotu.oscill}
            \Psi\sim e^{\ii\pazocal{S}}\,,
        \end{align}
        $\pazocal{S}$ denoting the classical action.
        By applying the Wentzel--Kramers--Brillouin (WKB) approximation and expanding the action in powers of the Planck mass, one finds that the wave function becomes oscillatory around configurations satisfying the Hamilton--Jacobi equation. See Ref. \cite{capozziello.odintsov.hamiltonian} for details. In this case, the conjugate momenta are
        \begin{align}\label{hamilton.jacobi.superspace}
            \pi^{ij} = \frac{\delta \pazocal{S}}{\delta h_{ij}}\,,
        \end{align}
        where $h_{ij}$ denotes the three-metric on spatial hypersurfaces. Eq.\eqref{hamilton.jacobi.superspace} corresponds to classical trajectories, thereby recovering classical spacetimes as discussed in \cite{wiltshire.2003.QC}. 

        In a minisuperspace context, the functional derivatives reduce to ordinary derivatives. Moreover, Noether's theorem ensures that, in the presence of symmetries, the Wave Function of the Universe exhibits  oscillatory behaviors, an indication that  cosmological observables are correlated and then classical universes can emerge 
        \cite{Capozziello:2022vyd}. In this framework, the classical evolution of  minisuperspace variables $q^i$ can be derived from 
        \begin{align}\label{hamilton.jacobi.minisuperspace}
            \pi_i = \frac{\partial \pazocal{S}}{\partial q^i}\,.
        \end{align}
       
An important remark is necessary at this point. The Noether Symmetry Approach is used as a selection rule to determine specific functional forms of the gravitational Lagrangian. As  discussed in details in Ref. \cite{bajardi.book.noether}, the  physical interpretation  emerges directly from the mathematical consistency. In fact, it is possible to state that any Noether symmetry has a physical interpretation because it corresponds to some conserved \qm{charge}. Several examples are reported in literature related to this statement. For example, in \cite{Capozziello:2007wc}, it is shown that the Noether symmetry is a function of the gravitational radius of spherically symmetric solutions in $f(R)$ gravity.   In particular, for $f(R)=R$, the Schwarzschild radius is recovered. In the present case,   the selected power-law forms of $f(R)$, $f(Q)$, and $f(T)$ should be regarded as physically preferred models because give rise to physical solutions. In quantum cosmology, the existence of a Noether symmetry means that the Hartle criterion is valid \cite{Capozziello:1999xr} and then the related solutions give rise to \qm{observable universes}. Clearly also models without symmetries can give rise to observable models but this not a limitation for the validity of the present method.

Another important point to stress is the fact that  the minisuperspace approach is only an approximation where the Hartle criterion indicates classical universes when   oscillatory wave functions are selected. However, in this framework, one is not dealing with the full theory, but just with a truncation  based on a finite number of degrees of freedom.  The final goal should be to recover  equivalence not only at the minisuperspace level but also within the full  (non-truncated) quantum-gravity setting. In this case, the number of degrees of freedom is infinite and a functional analysis approach must be developed for the whole superspace. Nevertheless, this task goes beyond the scope of the present paper.
 
 With this approach in mind, let us now proceed to the discussion of minisuperspaces emerging from Extended Trinity Gravity.

\section{Cosmology in Extended General Relativity} \label{section.fR}

    We begin our investigation with $f(\lc{R})$ gravity in metric formalism \cite{Capozziello.ext.gr.2011}, where the action is written in terms of a generic function of the Ricci scalar, namely
    \begin{align}
        \pazocal{S}=\int\dd^4x\sqrt{-g}f(\lc{R})\,,
    \end{align}
    and reduces to the standard Einstein-Hilbert action  for $f(\lc{R})=\lc{R}$. The connection, as in GR, is Levi-Civita as in Eq.\eqref{lc.connection}. Strictly speaking, every term involving the connection, such as covariant derivatives, curvature tensors or curvature scalars, have to be denoted with the symbol \qm{$\,\circ\,$}, as in \qm{$\,\lc{R}\,$}. However, since in GR it is $\tensor{\Gamma}{^\alpha_\mu_\nu}=\tensor{\lc{\Gamma}}{^\alpha_\mu_\nu}$, in this section, we omit the \qm{$\,\circ\,$} superscript on connection-related quantities.

    Let us take into account  now  a spatially flat Friedmann--Lema\^itre--Robertson--Walker (FLRW) geometry for our minisuperspace approach. The  metric reads as 
    \begin{align} \label{flrw.metric}
        \dd s^2 = \dd t^2 - a^2(t)[\dd x^2+\dd y^2 + \dd z^2]\,,
    \end{align} 
    with $a$  the  scale factor. The Ricci scalar  takes the form
    \begin{align}\label{ricci.scalar.value}
        R = - 6 \left( \frac{\ddot{a}}{a} + \frac{\dot{a}^2}{a^2} \right)\,,
    \end{align}
    which shows that the action should in principle be viewed as a function of the scale factor only. Despite that, it is convenient to treat $a$ and $R$ as two independent variables, and then to constrain $R$  via a Lagrange multiplier $\lambda$. Thus,  $\pazocal{Q}\defeq\{a, R\}$ is the configuration space, and the action can then be expressed as
    \begin{align}\label{f(R).action}
        \pazocal{S}=\int\dd t\,a^3\biggl[f(R)-\lambda\biggl(R+6\frac{\ddot{a}}{a}+6\frac{\dot a^2}{a^2}\biggr)\biggr]\,,
    \end{align}
    where  the cosmological time $t$ is the parameter that governs the evolution of the system. The variation of $\pazocal{S}$ with respect to the Ricci scalar yields  the value of the Lagrange multiplier
    \begin{align}
        \lambda=f_R(R)\,,
    \end{align}
    where hereafter  $f_R(R)\defeq \dd f(R)/ \dd R$. After integration by parts   \cite{Capozziello-DeFelice},  the Lagrangian can be written in the form
     \begin{align}
        \pazocal{L}=a^3\bigl(f(R)-Rf_R(R)\bigr)+6a\dot{a}^2f_R(R)+6a^2\dot a\dot R f_{RR}(R)\,.
    \end{align}
    The energy condition $E_\pazocal{L}=0$ (cf. Eq. \eqref{energy.from.lagrangian}) corresponds to the time-time component of the Einstein equations and  yields
    \begin{align}\label{EL.f(R)}
        6 a^2 \dot{a} \dot{R} f_{RR}(R) + 6 a \dot{a}^2 f_R(R) - a^3[f(R) - R f_R(R)] = 0\,,
    \end{align}
    while the Euler--Lagrange equations for $a$ and $R$ are
    \begin{subequations}
        \begin{align}
            &\dot{R}^2 f_{RRR}(R)  + \ddot{R} f_{RR}(R) + \frac{\dot{a}^2}{a^2} f_R(R) + 2 \frac{\ddot{a}}{a} f_R(R)  \notag
            \\
            &- \frac{1}{2} [f(R) - R f_R(R)] + 2\frac{\dot{a}}{a} \dot{R} f_{RR}(R) = 0\,,
            \\
            &R = - 6 \left( \frac{\ddot{a}}{a}+  \frac{\dot{a}^2}{a^2} \right)\,,
      \end{align}
    \end{subequations}
    respectively. 
    
    We resort to the Noether Symmetry Approach to reduce and solve the system. Thus, following  Sec. \ref{subsec.nsa}, we impose the condition \eqref{symmetry.definition} where $X$ is
    \begin{align}
        X=\alpha\frac{\del}{\del a} + \beta \frac{\del}{\del R} + \dot\alpha\frac{\del}{\del\dot a} + \dot\beta \frac{\del}{\del\dot R}\,,
    \end{align}
    and set all the independent coefficients of the resulting  expression to zero. In this way, we get the  equations
    \begin{subequations} \label{noether.system.fR}
        \begin{align}
            &f_R(\alpha + 2a\del_a\alpha) + af_{RR}(\beta+a\del_a\beta)=0\,, \\
            &a^2f_{RR}\del_R\alpha=0\,, \\
            &2f_R\del_R\alpha + f_{RR}(2\alpha + a\del_a\alpha + a\del_R\beta) + \alpha\beta f_{RRR}=0\,, \\
            &3\alpha(f-Rf_R) - a\beta Rf_{RR}=0\,.
        \end{align}
    \end{subequations}
    This system can be  solved analytically  \cite{Capozziello-DeFelice}. A possible solution is then
    \begin{subequations}
        \begin{align}
            &X=\frac{\alpha_0}{a}\del_a - 2\alpha_0\frac{R}{a^2}\del_R\,, \\
            &j_0=9\alpha_0f_0\biggl(R\dot{a}+\frac{1}{2}a\dot{R}\biggr)R^{-\frac{1}{2}}\,, \\
            &f(R)=f_0R^{3/2}\,,
        \end{align}
    \end{subequations}
    with $\alpha_0$ being an integration constant.
    
    We can reduce the system by transforming the minisuperspace variables from $\{a, R\}$ to $\{z, w\}$ in such a way that Eqs.\eqref{reduction.condition} are satisfied. We thus arrive at 
    \begin{subequations}
        \begin{align}
            &\alpha\del_az(a, R) + \beta\del_Rz(a,R)=1\,, \\
            &\alpha\del_aw(a, R) + \beta\del_Rw(a,R)=0\,,
        \end{align}
    \end{subequations}
    whose possible solution is 
    \begin{subequations}
        \begin{align}
            &\displaystyle z=\frac{a^2}{2\alpha_0}\,, \\
            &\displaystyle w=w_0\bigl(a\sqrt{R}\,\bigr)^\ell\,,
        \end{align}
    \end{subequations}
    where $w_0$ and $\ell$ are integration constants.  Writing the Lagrangian in terms of $z$ and $w$, we get
    \begin{align}\label{fR.lagrangian}
        \pazocal{L}(z,w)=\frac{f_0}{\ell w}\biggl(\frac{w}{w_0}\biggr)^{\frac{1}{\ell}} \Biggl[ 18\alpha_0\dot w\dot z -w\ell\biggl(\frac{w}{w_0}\biggr)^{\frac{2}{\ell}} \Biggr]\,,
    \end{align}
    where, as expected,  $z$ is a cyclic variable. The equations of motion in the new coordinates become
    \begin{subequations}
        \begin{align}
            &\displaystyle (\ell-1)\dot w^2-\ell w\ddot w=0\,, \\
            &\displaystyle \biggl(\frac{w}{w_0}\biggr)^{\frac{2}{\ell}} + 6 \alpha_0\ddot z=0\,
        \end{align}
    \end{subequations}
    and can be solved analytically. A  solution is 
    \begin{subequations}
        \begin{align}
            \displaystyle w &= w_0(t-t_0)^\ell\,, \\
            \displaystyle z& = z_0+z_1t+z_2t^2+z_3t^3+z_4t^4\,,
        \end{align}
    \end{subequations}
    with $\{z_i\}_{i=0}^4$ being integration constants. Switching back to the old variables, we finally obtain the evolution of the scale factor, which can be written as
    \begin{align}\label{f.R.scale.factor}
        a(t) = a_0 \sqrt{c_0+c_1t+c_2t^2+c_3t^3+c_4t^4}\,.
    \end{align}
    An  interesting aspect of \eqref{f.R.scale.factor} is that it can reproduce smooth transitions between different cosmological epochs as discussed in \cite{Moruno.a.fR, Rubano.a.fR}.

    An alternative route to achieve classical observables is  provided by minisuperspace quantum cosmology. Performing a Legendre transformation on the Lagrangian \eqref{fR.lagrangian}, we obtain the Hamiltonian 
    \begin{align}
        \pazocal{H} &= \pi_w\dot w+\pi_z\dot z -\pazocal{L} \nn\\
        &= \frac{\ell w}{18\alpha_0f_0} \biggl(\frac{w}{w_0}\biggr)^{-\frac{1}{\ell}}\pi_w\pi_z+f_0\biggl(\frac{w}{w_0}\biggr)^{\frac{3}{\ell}}\,.
    \end{align}
    The quantization of the system is achieved by applying the canonical substitution $\pi_i\rightarrow-\ii\del_i$, which leads to the Wheeler--DeWitt equation \cite{DeWitt-1967} for the minisuperspace  \cite{Capozziello:2022}
    \begin{align}\label{fr.wdw}
        \Biggl[-\frac{\ell w}{18\alpha_0f_0} \biggl(\frac{w}{w_0}\biggr)^{-\frac{1}{\ell}}\del_w\del_z+f_0\biggl(\frac{w}{w_0}\biggr)^{\frac{3}{\ell}}\Biggr]\Psi(z,w)=0\,.
    \end{align}
    Since $z$ is cyclic, its associated momentum is a constant of motion, and the corresponding quantum equation reads as
    \begin{align}\label{fr.quantum.momentum}
        \pi_z\Psi=-\ii\del_z\Psi=j_0\Psi\,.
    \end{align}
    We can then integrate \eqref{fr.wdw} and \eqref{fr.quantum.momentum} independently by separation of variables, obtaining the solution  \cite{Capozziello:2022}
    \begin{align}\label{fR.wavefunction}
        \Psi(z,w)= \Psi_0\exp\Biggl\{\ii\Biggl[j_0z-\frac{9\alpha_0f_0^2}{2j_0}\biggl(\frac{w}{w_0}\biggr)^{\frac{4}{\ell}}\Biggr]\Biggr\}\,.
    \end{align}

   Bearing in mind that in the semiclassical region the wave function can be approximated as \eqref{wfotu.oscill}, then it follows, from  Eq.\eqref{fR.wavefunction},  that the action is
    \begin{align}
        \pazocal{S}=j_0z-\frac{9\alpha_0f_0^2}{2j_0}\biggl(\frac{w}{w_0}\biggr)^{\frac{4}{\ell}}\,.
    \end{align}
    According to the WKB approximation, it is
    \begin{subequations}\label{fR.hj}
        \begin{align}
            &\pi_z=\frac{\del\pazocal{S}}{\del z}\,,\\
            &\pi_w=\frac{\del\pazocal{S}}{\del w}\,,
        \end{align}
    \end{subequations}
    whose solution, after inverting the variables,  leads to the previosly found scale factor
     \eqref{f.R.scale.factor}. It is worth saying that, according to the Hartle criterion, the oscillatory behavior of the wave function gives rise to an observable universe.     
     
     It is important to stress that this one of the possible solutions for the $f(R)$ minisuperspace cosmology \cite{Capozziello:2022}. However, as we will see below, it is an indicative case which can be confronted with solutions coming from extended STEGR and TEGR. 

\section{Cosmology in Extended Symmetric Teleparallel Gravity} \label{section.fQ.fQB}

    Let us deal now with $f(Q)$ gravity models, which provide  extensions of STEGR.
    
    Within STEGR, the connection $\tensor{\nm{\Gamma}}{^\alpha_\mu_\nu}$ is non-metric, while being curvature-free and symmetric. To be rigorous, we should mark all the connection related quantities with the superscript \qm{$\diamond$} but, unless there is some risk of ambiguity, we will omit it for the sake of simplicity. The gravitational interaction is encoded in the non-metricity tensor \eqref{non.metricity}, and 
    the  action can be expressed as
    \begin{align}\label{stegr.action}
        \pazocal{S}=\int\dd^4x\sqrt{-g}Q\,,
    \end{align}
    where   
    \begin{align}
        Q\defeq\frac{1}{4} Q_\alpha Q^\alpha -\frac{1}{4} Q_{\alpha\beta\gamma} Q^{\alpha\beta\gamma} + \frac{1}{2} Q_{\alpha\beta\gamma} Q^{\beta\alpha\gamma}-\frac{1}{2} Q_\alpha  {\bar{Q}}^\alpha\,,
    \end{align}
   denotes the \textit{non-metricity scalar},  with
    \begin{subequations}
        \begin{align}
            &Q_\alpha\defeq g^{\mu\nu}Q_{\alpha\mu\nu}\,, \\
            &\bar{Q}_{\alpha}\defeq g^{\mu\nu}Q_{\mu\nu\alpha}\,.
        \end{align}
    \end{subequations}
    The scalar $Q$ satisfies the identity
    \begin{align}\label{R.Q.relation} 
        \lc{R}=Q-B\,,
    \end{align}
    with $\lc{R}$ the Ricci curvature scalar associated to the Levi-Civita connection, and 
    \begin{align}\label{fQ.divergence.term}
        B=\frac{1}{\sqrt{-g}}\del_\alpha\bigl[\sqrt{-g}(Q_\alpha-\bar{Q}_\alpha)\bigr].
    \end{align}
    a boundary term.
    Since $B$ is a total divergence, it does not affect the field equations and can be neglected in the action. Consequently, the action \eqref{stegr.action} is dynamically equivalent to the Einstein-Hilbert action, thereby justifying the designation of \qm{STEGR}.  
    
    However, this correspondence no longer holds for generic STEGR extensions in which \eqref{stegr.action} is generalized to include an arbitrary function $f(Q)$ of the non-metricity scalar.  As we will see  in Sec. \ref{subsec.fQB}, the equivalence with $f(R)$ gravity can be restored if the Lagrangian belongs to the specific subclass $f(Q-B)$ within the broader family of $f(Q,B)$ models.

    \subsection{\texorpdfstring{$f(Q)$ cosmology}{f(Q) cosmology}}\label{subsec.fQ}

        Following the procedure adopted in Sec. \ref{section.fR}, we start from the   action \cite{Heisenberg_fQ}
         \begin{align}
            \pazocal{S}=\int \dd^4x\sqrt{-g}f(Q)\,,
        \end{align}
        and derive the action for a flat FLRW universe:
         \begin{align}\label{fQ.flrw.action}
            \pazocal{S}=\int \dd t a^3\Biggl[ f(Q)-\lambda\biggl( Q - 6\frac{\dot{a}^2}{a^2} \biggr) \Biggr]\,
        \end{align}
        where variation with respect to the non-metricity scalar fixes the value of the Lagrange multiplier $\lambda$ to
        \begin{align}\label{fQ.lagrange.mult}
            \lambda=f_Q (Q)\,.
        \end{align}
        The point-like Lagrangian  takes the form
        \begin{align} \label{initial.Lagrangian.fQ}
            \pazocal{L}=a^3(f-Qf_Q)+6f_Qa\dot{a}^2\,,
        \end{align}
        which allows to evaluate the Euler--Lagrange equations along with the energy condition $E_{\pazocal{L}}=0$, thus yielding 
        \begin{subequations}
            \begin{align}
                &4f_Qa\ddot{a} + 2f_Q\dot{a}^2 + 4f_{QQ}\dot{a}\dot{Q} + a^2(f-Qf_Q)=0\,, \label{fQ.eom.a}\\
                &Q=6\frac{\dot{a}^2}{a^2}\,,\label{fQ.eom.Q}\\
                &6f_Q\dot{a}^2-a^2(f-Qf_Q)=0\,.\label{fQ.ec}
            \end{align}
        \end{subequations}
        By substituting \eqref{fQ.eom.Q} into \eqref{fQ.eom.a} and \eqref{fQ.ec}, we obtain \cite{Bajardi-2020-bouncing, Bajardi-2023}      
        \begin{subequations}
            \begin{align}
                &12f_Q\dot{a}^2-a^2f=0\,,\\
                &12f_{QQ}\dot{a}^2+a^2f_Q=0\,.
            \end{align}
        \end{subequations}
         According to what we have done for  $f(R)$ gravity, we seek for a symmetry  that permits to reduce the dynamical system. The symmetry generator is
        \begin{align}
            X=\,&\alpha(a,Q)\frac{\del}{\del a} + \beta(a,Q) \frac{\del}{\del Q} + \dot\alpha(a,Q)\frac{\del}{\del\dot a} \nn \\
            & + \dot\beta(a,Q) \frac{\del}{\del\dot Q}\,.
        \end{align}
       From identity  \eqref{symmetry.definition},   we require that all independent coefficients  vanish, thus  obtaining 
        \begin{subequations}
            \begin{align}
                & \alpha f_Q + a\beta f_{QQ} + 2af_Q\del_a\alpha=0\,, \label{noether.fQ.eqnA}\\
                & af_Q\del_Q\alpha=0\,, \label{noether.fQ.eqnB}\\
                & 3\alpha(f-Qf_Q)-a\beta Qf_{QQ}=0\,. \label{noether.fQ.eqnC}
            \end{align}
        \end{subequations}
        From \eqref{noether.fQ.eqnB}, we derive $\alpha=\alpha(a)$, whereas a dimensional analysis of the remaining two equations suggests that $\beta$ has to be
        \begin{align}\label{beta.fQ.ansatz}
            \beta=c_1\frac{Q}{a}\alpha(a)+c_2Q\del_a\alpha\,,
        \end{align}
         $c_1$ and $c_2$ being arbitrary constants. By plugging \eqref{beta.fQ.ansatz}, together with the \emph{ansatz} $f(Q)=f_0Q^s$ ($s \in \mathbb{R}$), into \eqref{noether.fQ.eqnA} and \eqref{noether.fQ.eqnC}, we can find the value of the coefficients of the generator $X$,  and  determine $s$. Discarding the STEGR case $s=1$, we get
        \begin{subequations}
            \begin{align}
                &X=\alpha_0a^{\frac{2+c_1}{2-c_2}}\del_a + 2\alpha_0\frac{c_1+c_2}{2-c_2}Qa^{\frac{c_1+c_2}{2-c_2}}\del_Q\,, \\
                &j_0=18f_0\frac{c_2-2}{c_1+c_2}a^{\frac{4+c_1-c_2}{2-c_2}}Q^{\frac{c_2-2c_1-6}{2c_1+2c_2}}\dot{a}\,, \\
                &f(Q)=f_0Q^{-\frac{3}{2} \left(\frac{2-c_2}{c_1+c_2}\right)}\,.
            \end{align}
        \end{subequations}
        To simplify these expressions and to explicitly show the dependence of the Noether vector field on the underlying  gravity model, we  express $c_1$ and $c_2$ in terms of parameter $s$ as follows:
        \begin{align}
            s=-\frac{3}{2}\left(\frac{2-c_2}{c_1+c_2}\right)\,,
        \end{align}
        so that
        \begin{subequations}
            \begin{align}
                &X=\alpha_0a^{\frac{2s-3}{2s}}\del_a - \frac{3\alpha_0}{s}Qa^{-\frac{3}{2s}}\del_Q\,, \\
                &j_0=-12s\,\alpha_0f_0a^{\frac{4s-3}{2s}}Q^{s-1}\dot{a}\,, \\
                &f(Q)=f_0Q^{s}\,.
            \end{align}
        \end{subequations}
        Substituting the expression of $f(Q)$ into \eqref{initial.Lagrangian.fQ}, we get the Lagrangian in the $\{a,Q\}$ variables
        \begin{align}\label{lagrangian.fQ.Qs}
            \pazocal{L}=f_0(1-s)a^3Q^s+6f_0sQ^{s-1}a\dot{a}^2\,.
        \end{align}
        We can perform the change of coordinates $\{a,Q\}\rightarrow\{z,w\}$ to reduce the system. Imposing the condition \eqref{reduction.condition}, we derive a possible solution
        \begin{subequations}
            \begin{align}
                & z=\frac{2s}{3\alpha_0}a^{\frac{3}{2s}}\,, \\
                & w=a^3Q^s\,,
            \end{align}
        \end{subequations}
        which, when  inserted into  \eqref{lagrangian.fQ.Qs}, yields  the Lagrangian in the new coordinates
        \begin{align}\label{fQ.lagrangian.zw}
            \pazocal{L}=f_0w(1-s)+6f_0s\alpha_0^2w^{\frac{s-1}{s}}\dot{z}^2
        \end{align}
        and the associated Euler--Lagrange equations
        \begin{subequations}\label{fQ-zw-el-eqns}
            \begin{align}
                &sw\ddot{z}+(s-1)\dot{w}\dot{z}=0\,, \label{fQ-z-el-eqn}\\
                &w=6^s\alpha_0^{2s}\dot{z}^{2s}\,. \label{fQ-w-el-eqn}
            \end{align}
        \end{subequations}
        These can be analytically solved to derive $z$, and subsequently $a$, thereby finding 
        \begin{align} \label{sol-f-Q-z-a}
            z(t)=z_0t\,,\quad a(t)=a_0t^{\frac{2s}{3}}\,.
        \end{align}
        If we  further impose the energy condition $E_\pazocal{L}=0$, which reads as
        \begin{align}
            6sf_0\alpha_0^2w^{\frac{s-1}{s}}\dot{z}^2-f_0(1-s)w=0\,,
        \end{align}
        then, using  \eqref{fQ-w-el-eqn}, we get $s=1/2$. This value gives the scale factor
        \begin{align} \label{fQ.scale.factor}
            a(t)\sim t^{1/3}\,,
        \end{align}
        which corresponds to a stiff matter-dominated universe.

        We can now perform a Legendre transform of the Lagrangian \eqref{fQ.lagrangian.zw} to derive the Hamiltonian and  quantize the system.
        An interesting aspect of the $f(Q)$ setup is that, upon  labeling the set of coordinates $\{z,w\}$ with $q^i$, it is straightforward to verify that the Lagrangian does not satisfy condition \eqref{non.singularity.condition}, indicating that we are dealing with a constrained Hamiltonian system.  A similar model was studied in  the context of  $f(T)$ gravity in \cite{Capozziello:2022,Capozziello:2022vyd}. 
        
        Starting from Eq. \eqref{fQ.lagrangian.zw},  we compute the canonical momenta
        \begin{align}
            \pi_z=12s f_0 \alpha_0^2 w^{\frac{s-1}{s}}\dot{z}
        \end{align}
        and the primary constraint
        \begin{align}
            \varphi_1\defeq\pi_w\approx0\,,
            \label{constraint-f-Q-Dirac}
        \end{align}
        with \qm{$\approx$} representing an equality holding on the constraint submanifold \cite{Esposito-QG-QC}. Therefore, we are led to the  Hamiltonian
        \begin{align}
            \tilde{\pazocal{H}}=\pazocal{H}+\mu \varphi_1\,,
        \end{align}
        with $\pazocal{H}$  the canonical Hamiltonian
        \begin{align}\label{fQ.canonical.Hamiltonian}
            \pazocal{H}=\pi_z\dot{z}-\pazocal{L}=\frac{\pi_z^2}{24s f_0 \alpha_0^2}w^{\frac{1-s}{s}}+f_0w(s-1)\,,
        \end{align}
        and $\mu$  the Lagrange multiplier for the constraint \eqref{constraint-f-Q-Dirac}. The equations of motion thus reads as 
        \begin{subequations}
            \begin{align}
                &\dot{z}=\frac{\del\tilde{\pazocal{H}}}{\del\pi_z}=\frac{\pi_z}{12s f_0 \alpha_0^2 w^{\frac{s-1}{s}}}\,, \\
                &\dot{w}=\mu\,,\\
                &\dot{\pi}_z=0\,, \label{fQ.pz.conservation}\\
                &\dot{\pi}_w=-\frac{\pi_z^2(1-s)}{24s^2f_0\alpha_0^2}w^{\frac{1-2s}{s}}-f_0(s-1)\,.
            \end{align}
        \end{subequations}
        Asking for the primary constraint $\varphi_1$ be preserved under time evolution, i.e., $\dot{\varphi}_1\approx0$ leads to the secondary constraint
        \begin{align}
            \varphi_2\defeq-\frac{\pi_z^2(1-s)}{24s^2f_0\alpha_0^2}w^{\frac{1-2s}{s}}-f_0(s-1)\approx0\,,
        \end{align}
        which is equivalent to the Euler--Lagrange equation \eqref{fQ-w-el-eqn}, while the remaining Hamilton equations lead to \eqref{fQ-z-el-eqn}. Finally, we can impose the consistency condition $\dot{\varphi}_2\approx0$, which yields
        \begin{align}
            \{\varphi_2,\tilde{\pazocal{H}}\} =\frac{\pi_z^2(s-1)(2s-1)}{24s^3f_0\alpha_0^2}\mu w^{\frac{1-3s}{s}}\approx0\,,
        \end{align}
        with $\{\,,\,\}$ being the ordinary Poisson brackets. We thus see that this condition is verified if one of the following  possibilities occurs:
        \begin{itemize}
            \item $\mu=0$, which can be readily discarded, as this choice  makes the Hamiltonian pathological again; 
            \item $s=1$, which has been excluded because it is the trivial case of  STEGR;
            \item $s=1/2$, in agreement with the Lagrangian formulation.
        \end{itemize}

        To quantize the minisuperspace, we perform the standard canonical substitutions $\pi_i\rightarrow-\ii\del_i$ in the energy condition $E_{\pazocal{L}}=\tilde{\pazocal{H}}=0$ and in Eqs.  \eqref{constraint-f-Q-Dirac} and \eqref{fQ.pz.conservation}. In this way, we find that the equations that fully determine the state of the system read as
        \begin{subequations}
            \begin{align}
                &\tilde{\pazocal{H}}\Psi(z,w)=0\,,\label{fQ.wdw}\\
                &-\ii\del_z\Psi(z,w)=j_0\Psi(z,w)\,,\label{fQ.quantum.conservation}\\
                &-\ii\del_w\Psi(z,w)=0\,, \label{fQ.quantum.constraint}
            \end{align}
        \end{subequations}
        which represent the Wheeler--DeWitt equation for the minisuperspace, the conservation of $\pi_z$, and the primary constraint condition, respectively,

       We first exploit \eqref{fQ.quantum.constraint} to eliminate the $w$ dependence and  then, by  using \eqref{fQ.quantum.conservation},  we find that   $\Psi$ assumes the form
        \begin{align}
            \Psi(z)=\Psi_0e^{\ii j_0z}\,.
        \end{align}
        Substituting this expression into \eqref{fQ.wdw} yields
        \begin{align}
            j_0^2=24s(1-s)f_0^2\alpha_0^2w^{\frac{2s-1}{s}}\,,
        \end{align}
        so that the sought-after  wave function can be written as
        \begin{align}\label{fQ.wavefunction.universe}
            \Psi(z)=\Psi_0\exp\Bigl\{\ii\bigl[2f_0\alpha_0w^{\frac{2s-1}{2s}}\sqrt{6s(1-s)}\bigr]z\Bigr\}\,,
        \end{align}
        which implies that the classical action  is (cf. Eq. \eqref{wfotu.oscill})
        \begin{align}
            \pazocal{S}=2f_0\alpha_0w^{\frac{2s-1}{2s}}\sqrt{6s(1-s)}\,z\,.
        \end{align}
        Employing the Hartle criterion stated in Sec. \ref{subsec.nsa}, we need to consider the Hamilton--Jacobi equations \eqref{hamilton.jacobi.minisuperspace}, which lead to 
        \begin{align}
            & w^{\frac{1}{2s}}=\sqrt{\frac{6s}{1-s}}\alpha_0\dot{z}\,, \label{semiclassical-z-fQ}
            \\
            &   2f_0\alpha_0z\biggl( \frac{2s-1}{2s} \biggr)w^{-\frac{1}{2s}}=0\,, \label{semiclassical-w-fQ}
        \end{align}
        the latter being satisfied for $s=1/2$\,. By substituting  this value  into \eqref{semiclassical-z-fQ}, we obtain
        \begin{align}\label{fq.hamiltonjacobi.w}
            w=\sqrt{6}\alpha_0\dot{z}\,,
        \end{align}
        which is equivalent to \eqref{fQ-w-el-eqn}.

        A last remark is now in order. We notice that we quantized $\tilde{\pazocal{H}}$, along with the constraint $\varphi_1\Psi=0$, without considering neither the total Hamiltonian
        \begin{align}
            \pazocal{H}_T\defeq\tilde{\pazocal{H}}+\mu_2\varphi_2\,,
        \end{align}
        nor the constraint condition
        \begin{align}
            \varphi_2\Psi=0\,.
        \end{align}
        Despite this procedure would have been more rigorous, employing it is not strictly necessary. The Wheeler--DeWitt equation, expressed in terms of $\pazocal{H}_T$, yields an equivalent result, while the constraint equation can be interpreted as an additional consistency condition that  gives again  $s = 1/2$.
        It is worth noticing that $f(Q)$ gravity does not give rise to solutions equivalent to $f(R)$ gravity. In order to restore equivalence, we need to introduce a boundary term. 

    \subsection{\texorpdfstring{$f(Q,\hat{B})$ cosmology}{f(Q,B) cosmology}} \label{subsec.fQB}
    
        In the previous section, we  showed that $f(Q)$  and $f(R)$ gravity are not equivalent. As noted before \cite{Capozziello-ext-trinity-2025}, the correspondence can be recovered by considering a Lagrangian of the form $f(Q-B)$. We now prove that this result holds also at the minisuperspace level, by first developing a cosmological model for the most general $f(Q,B)$ scenario and then specializing to the $f(Q-B)$ case.
    
        By adopting the minisuperspace $\pazocal{Q}=\{a, Q, B\}$, the action pertaining to a flat FLRW universe is 
        \begin{align}\label{fQB.action}
            \pazocal{S}=&\int\dd t \,  a^3 \Biggl[f(Q,B)-\lambda\biggl(Q - 6\frac{\dot a^2}{a^2}\biggr) \nn\\
            &- \rho\biggl( B - 6\frac{\ddot a}{a} -12 \frac{\dot a^2}{a^2}\biggr)\Biggr]\,.
        \end{align}
        By performing the variations of $\pazocal{S}$ with respect to $Q$ and $B$, we get the Lagrange multipliers
        \begin{subequations}\label{fQB.lagrange.multipliers}
            \begin{align}
                \lambda &= \del_Qf(Q,B)\defeq f_Q, \\
                \rho &= \del_Bf(Q,B)\defeq f_B\,.
            \end{align}
        \end{subequations}
        As a preliminary consistency check, we verify that  the equivalence between $f(Q,B)$ and $f(R)$ is restored for the special case $f(Q, B)=f(Q-B)$. This can be easily seen  if we exploit
        \begin{align}
            &\frac{\del}{\del Q}= \frac{\del}{\del(Q-B)}\,, \\
            &\frac{\del}{\del B}= - \frac{\del}{\del(Q-B)} \,,
        \end{align}
        then the action becomes
        \begin{align}\label{f(Q-B).action}
            \pazocal{S} =& \int\dd t \, a^3 \Biggl[f(Q-B) \nn \\
            &-f_{Q-B}\biggl(Q-B + 6\frac{\ddot a}{a}+6\frac{\dot a^2}{a^2}\biggr)\Biggr]\,,
        \end{align}
        which is equivalent to \eqref{f(R).action} according to \eqref{R.Q.relation}. While one might argue that this equivalence is merely \textit{formal}, we note that the same Lagrange multiplier fixes the value of both $\lc{R}$  and  $Q-B$  to be equal to the Ricci  curvature scalar in both $f(R)$ and $f(Q,B)$ gravity. Therefore, the two actions are equivalent. 
    
        Let us now return to  the general $f(Q,B)$ gravity setup.  It follows from \eqref{fQB.action} and \eqref{fQB.lagrange.multipliers} that the Lagrangian function reads as
        \begin{align}\label{fQB.lagrangian}
            \pazocal{L}=&a^3f-a^3Qf_Q-a^3Bf_B \nn \\
            &+ 6af_Q\dot{a}^2-6a^2f_{BQ}\dot a\dot Q-6a^2f_{BB}\dot a \dot B\,,
        \end{align}
        so that the Euler--Lagrange equations for $a$, $Q$ and $B$ are 
        \begin{subequations}\label{eom.fQB}
            \begin{align}
                &f_Q\dot{a}^2+2f_Qa\ddot{a} +2f_{QQ}a\dot{a}\dot{Q} +2f_{QB}a\dot{a}\dot{B} -2f_{BBQ}a^2\dot{Q}\dot{B} \nn \\ 
                &-f_{BQQ}a^2\dot{Q}^2 -f_{BQ}a^2\ddot{Q} -f_{BBB}a^2\dot{B}^2 -f_{BB}a^2\ddot{B} \nn \\
                & -\frac{a^2}{2}\bigl(f-Qf_Q-Bf_B\bigr)=0\,, \\
                &f_{BQ}\bigl(6a\ddot{a}-a^2B+12\dot{a}^2\bigr)+f_{QQ}\bigl(6\dot{a}^2-a^2Q\bigr)=0\,, \label{eom.fQB-2} \\
                &f_{BQ}\bigl(6\dot{a}^2-a^2Q\bigr)+f_{BB}\bigl(6a\ddot{a}-a^2B+12\dot{a}^2\bigr)=0\,, \label{eom.fQB-3}
            \end{align}
        \end{subequations}
        respectively. 
        For $f_{QQ}$, $f_{QB}$, and $f_{BB}$ different from zero, \eqref{eom.fQB-2} and \eqref{eom.fQB-3}  yield the equations for the constraints 
        \begin{align}
            Q&=6\frac{\dot{a}^2}{a^2}\,, \\
            B&=6\frac{\ddot{a}}{a}+12\frac{\dot{a}^2}{a^2}\,. 
        \end{align}
        We can now supplement the equations of motion \eqref{eom.fQB} with the energy condition
        \begin{align}\label{fQB.energy.condition}
            E_\pazocal{L}=&6f_Qa\dot{a}^2 -6f_{BQ}a^2\dot{a}\dot{Q} \nn \\
            &-6f_{BB}a^2\dot{a}\dot{B} -a^3\bigl(f-Qf_Q-Bf_B\bigr)=0\,,
        \end{align}
        which completes the description of the dynamical system. Since the set of Eqs. \eqref{eom.fQB} and \eqref{fQB.energy.condition} cannot be solved analytically, we reduce  the dynamical system by adopting the Noether Symmetry Approach. By imposing \eqref{symmetry.definition} for the generator
        \begin{align}\label{fQB.generator}
            X=&\alpha(a,Q,B)\frac{\del}{\del a}+\beta(a,Q,B)\frac{\del}{\del Q}+\gamma(a,Q,B)\frac{\del}{\del B} \nonumber\\
            &+\dot{\alpha}(a,Q,B)\frac{\del}{\del\dot{a}}+\dot{\beta}(a,Q,B)\frac{\del}{\del\dot{Q}}+\dot{\gamma}(a,Q,B)\frac{\del}{\del\dot{B}}\,,
        \end{align}
        and by setting to zero all the independent terms in the ensuing calculations, we get the following set of partial differential equations (PDEs):
        \begin{subequations}\label{noether.fQB.equations}
            \begin{align}
                &f_Q\alpha +af_{QQ}\beta +af_{BQ}\gamma +2af_Q\del_a\alpha \nn \\
                &-a^2f_{BQ}\del_a\beta -a^2f_{BB}\del_a\gamma =0\,, \label{noether.fQB.eqnA}\\
                &f_{BQ}\del_Q\alpha=0\,, \label{noether.fQB.eqnB}\\
                &f_{BB}\del_B\alpha=0\,, \label{noether.fQB.eqnC}\\
                &f_{BQ}\bigl(2\alpha+a\del_a\alpha\bigr) +a\del_Q\bigl(f_{BQ}\beta+f_{BB}\gamma\bigr)=0\,, \label{noether.fQB.eqnD}\\
                &f_{BB}\bigl(2\alpha+a\del_a\alpha\bigr) +a\del_B\bigl(f_{BQ}\beta+f_{BB}\gamma\bigr)=0\,, \label{noether.fQB.eqnE}\\
                &f_{BQ}\del_B\alpha+f_{BB}\del_Q\alpha=0\,, \label{noether.fQB.eqnF}\\
                &3\alpha\bigl(f-Qf_Q-Bf_B\bigr)\alpha -af_{QQ}Q\beta \nn \\
                &-af_{BQ}B\beta -af_{BQ}Q\gamma-af_{BB}B\gamma=0\,.\label{noether.fQB.eqnG}
            \end{align}
        \end{subequations}
    
        To solve the system, we begin by considering \eqref{noether.fQB.eqnB} and \eqref{noether.fQB.eqnC}. We will suppose that $f_{QB}\neq0$ and $f_{BB}\neq0$, because otherwise we  get 
        \begin{align}
            f(Q,B)=A_1B+f(Q)\,,
        \end{align}
        with $A_1$  an integration constant and $f(Q)$  a generic function of $Q$  which, clearly, is equivalent to the  case already  studied in Sec. \ref{subsec.fQ}. Therefore, we can conclude that \eqref{noether.fQB.eqnB} and \eqref{noether.fQB.eqnC} imply 
        \begin{align}
            \del_Q\alpha=\del_B\alpha=0\,.
        \end{align}
        In order to find an analytical solution to the system \eqref{noether.fQB.equations}, we introduce  the following \emph{ansatz}:
        \begin{align}\label{f.QB.ansatz}
            f(Q,B)=f(Q+B/k)\,,
        \end{align}
        with $k\in\R$ being a dimensionless constant. We  then assume that all the dynamical variables, including $\alpha$, $\beta$ and $\gamma$, depend on the combination $Q+B/k$, rather than on $Q$ and $B$ separately. By setting
        \begin{align}
            &\tilde{R}\defeq Q+\frac{B}{k}\,, \\
            &\tilde{\beta}\defeq \beta+\frac{\gamma}{k}\,,
        \end{align}
       the Noether vector field \eqref{fQB.generator} becomes
        \begin{align}
            X=\alpha(a)\del_a+\tilde{\beta}(a,\tilde{R})\del_{\tilde{R}}+\dot{\alpha}(a)\del_{\dot{a}}+\dot{\tilde{\beta}}(a,\tilde{R})\del_{\dot{\tilde{R}}}
        \end{align}
        and the remaining equations \eqref{noether.fQB.eqnA} and \eqref{noether.fQB.eqnD}--\eqref{noether.fQB.eqnG} can be thus expressed in the form
        \begin{subequations}
            \begin{align}
                &\alpha f_{\tilde{R}} +a\tilde{\beta}f_{\tilde{R}\tilde{R}} +2a\del_a\alpha f_{\tilde{R}} -\frac{a^2}{k} f_{\tilde{R}\tilde{R}}\del_a\tilde{\beta}=0\,, \\
                &f_{\tilde{R}\tilde{R}}(2\alpha+a\del_a\alpha) +a\del_{\tilde{R}} \bigl(f_{\tilde{R}\tilde{R}}\tilde{\beta}\bigr) =0\,, \\
                &3\alpha\bigl(f-\tilde{R}f_{\tilde{R}}\bigr)-a\tilde{\beta}\tilde{R}f_{\tilde{R}\tilde{R}}=0 \,.
            \end{align}
        \end{subequations}
        Following the procedure  employed in the previous sections, we get the solution
        \begin{subequations}
            \begin{align}
                &X=\alpha_0a^{-(2+k)}\del_a -\alpha_0(3+k)\tilde{R}a^{-(3+k)}\del_{\tilde{R}}\,, \\
                &j_0=\frac{18\alpha_0f_0}{3+k}\biggl(\tilde{R}\dot{a}+\frac{a\dot{\tilde{R}}}{3+k}\biggr)a^{-(k+1)}\tilde{R}^{-\frac{3+2k}{3+k}}\,, \\
                &f(\tilde{R})=f_0\tilde{R}^{\frac{3}{3+k}}\,.
            \end{align}
        \end{subequations}
    
        We can now reduce the dynamics by performing the change of variables
        \begin{align}
            (a,\tilde{R})\,\,\longrightarrow\,\,(z,w), \label{change-variable-f-q-b-general}
        \end{align}
        where $z$ is cyclic. From \eqref{reduction.condition}, we get the solution
        \begin{subequations}
            \begin{align}
                &z=\frac{a^{3+k}}{\alpha_0(3+k)}\,, \\
                &w=w_0a^\ell\tilde{R}^{\frac{\ell}{3+k}}\,,
            \end{align}
        \end{subequations}
        while the Lagrangian \eqref{fQB.lagrangian} becomes
        \begin{align}\label{fQB.zw.lagrangian}
            \pazocal{L}=\frac{f_0}{(3+k)\ell w} \biggl(\frac{w}{w_0}\biggr)^{-\frac{k}{l}} \Biggl[18\alpha_0\dot{w}\dot{z} +k\ell w\biggl(\frac{w}{w_0}\biggr)^{\frac{3+k}{\ell}}\Biggr]\,.
        \end{align}
       Therefore, we readily obtain the equations of motion  
        \begin{subequations}
            \begin{align}
                &\ell w\ddot{w}-(k+\ell)\dot{w}^2=0 \,, \\
                &6\alpha_0\ddot{z}-k\biggl(\frac{w}{w_0}\biggr)^{\frac{3+k}{\ell}}=0\,,
            \end{align}
        \end{subequations}
        which permit  deriving the expression for $z(t)$:
        \begin{align}\label{z.fQB}
            z(t)=z_0+z'_0t-\frac{k^3}{18\alpha_0(k-3)}(t-t_0)^{\frac{k-3}{k}}\,,
        \end{align}
        or, equivalently, for the scale factor $a(t)$:
        \begin{align}\label{fQB.scalefactor}
            a(t)=\biggl[a_0+a'_0t -\frac{k^3(k+3)}{18(k-3)}(t-t_0)^{\frac{k-3}{k}}\biggl]^{\frac{1}{k+3}}\,,
        \end{align}
        where it is clear that  only for $k=-1$,   the solution \eqref{f.R.scale.factor} is recovered. Therefore,  at the minisuperspace level, a model defined  by \eqref{f.QB.ansatz} is not, in general, equivalent to  $f(R)$ gravity. In this perspective, we can say that $f(R)$ metric gravity is a particular case of the most general $f(Q,B)$ gravity.
    
        Performing a Legendre transformation of the Lagrangian \eqref{fQB.zw.lagrangian}, we obtain the Hamiltonian
        \begin{align}\label{fQB.zw.hamiltonian}
            \pazocal{H}&=\pi_z\dot{z}+\pi_w\dot{w}-\pazocal{L} \nn \\
            &= \frac{(3+k)lw}{18\alpha_0f_0}\biggl(\frac{w}{w_0}\biggr)^{\frac{k}{\ell}}\pi_z\pi_w-\frac{f_0k}{3+k}\biggl(\frac{w}{w_0}\biggr)^{\frac{3}{\ell}}\,,
        \end{align}
       which allows for the quantization of the dynamical system via the canonical substitutions $\pi_i\rightarrow-\ii\del_i$. Imposing the Hamiltonian constraint $\pazocal{H}=0$ yields the Wheeler--DeWitt equation for the minisuperspace
        \begin{align}\label{wheeler.dewitt.fQB}
            \Biggl[\frac{(3+k)lw}{18\alpha_0f_0}\biggl(\frac{w}{w_0}\biggr)^{\frac{k}{\ell}}\del_z\del_w +\frac{f_0k}{3+k}\biggl(\frac{w}{w_0}\biggr)^{\frac{3}{\ell}}\Biggr]\Psi(z,w)=0\,, 
        \end{align}
       which can be solved by separations of variables. Assuming
        \begin{align}
            \Psi(z,w)=\psi(z)\phi(w)
        \end{align}
        and substituting into \eqref{wheeler.dewitt.fQB}, we find the general solution
        \begin{align}\label{fQB.wavefunction}
            &\Psi(z,w)= \nn \\
            &\,\,\,\Psi_0\exp\Biggl\{\ii\Biggl[j_0z + \frac{18k\alpha_0f_0^2}{(3+k)^2(3-k)j_0}\biggl(\frac{w}{w_0}\biggr)^{\frac{3-k}{\ell}}\Biggr]\Biggr\}\,
        \end{align}
        and hence, from Eq.\eqref{wfotu.oscill},   the semiclassical  action
        \begin{align}
            \pazocal{S}=j_0z + \frac{18k\alpha_0f_0^2}{(3+k)^2(3-k)j_0}\biggl(\frac{w}{w_0}\biggr)^{\frac{3-k}{\ell}}\,.
        \end{align}
        According to the Hartle criterion, the Hamilton--Jacobi equations \eqref{hamilton.jacobi.minisuperspace} for the $f(Q+B/k)$ are satisfied. Since their solution gives  \eqref{z.fQB}, we can conclude that  the minisuperspace quantum  cosmology is consistent with  $f(Q,B)$ gravity. In particular, the  equivalence with the $f(R)$ gravity is restored  for $k = -1$.

\section{Cosmology in Extended  Teleparallel Gravity} \label{section.fT.fTB}

   Let us consider  now extensions of TEGR.  In this framework, gravity is treated as a gauge theory  for the translation group \cite{aldrovandi2012teleparallel}. In this setting, the connection  $\tensor{\hat{\Gamma}}{^\alpha_\mu_\nu}$ is  the \textit{Weitzenb\"ock connection}. It is metric-compatible and  curvature-free but non-symmetric. The gravitational interactions is fully encoded in the torsion tensor $\tensor{\hat{T}}{^\lambda_\mu_\nu}$ (cf. Eq. \eqref{torsion.comp}). For the sake of simplicity, we will drop the \qm{$_\text{\textasciicircum}$}, unless it is necessary. It can be  shown that, in the TEGR framework, the action takes the form  \cite{aldrovandi2012teleparallel}
    \begin{align}\label{tegr.action}
        S=-\int\dd^4x\,eT\,,
    \end{align}
    where $e=\det[\tensor{e}{^A_\mu}]=\sqrt{-g}$  denotes the determinant of the tetrad field $\tensor{e}{^A_\mu}$ (with $A=0, 1, 2, 3$  the tetrad internal indices and $\mu=0, 1, 2, 3$  the coordinate ones), and 
    \begin{align}
        T\defeq\frac{1}{4}\tensor{T}{^\lambda_\mu_\nu}\tensor{T}{_\lambda^\mu^\nu}+\frac{1}{2}\tensor{T}{^\lambda_\mu_\nu}\tensor{T}{^\nu^\mu_\lambda}-\tensor{T}{^\alpha_\mu_\alpha} \tensor{T}{^\alpha^\mu_\alpha}
    \end{align}
    the torsion scalar. Furthermore, the theory is characterized by the identity \cite{capozziello-defalco-ferrara}
    \begin{align}
        \lc{R}=-T-\hat{B}\,,
    \end{align}
    with
    \begin{align}\label{fT.divergence.term}
        \hat{B}\defeq\frac{2}{e}\del_\mu(eT^\mu)\,,
    \end{align}
    which represents the metric teleparallel analogue of the symmetric teleparallel Eq.\eqref{R.Q.relation}.  This equivalence allows us to derive the following conclusions: 
    \begin{itemize}
        \item up to a total divergence, the action \eqref{tegr.action} is equivalent to the Einstein-Hilbert one, just like \eqref{stegr.action}. This is why such theory is called TEGR, and together with GR and STEGR,  it constitutes the Geometric Trinity of Gravity;
        \item the equivalence breaks down for extended theories with torsion for the same reasons discussed for  $f(Q)$ and $f(Q,B)$ cases.
    \end{itemize}

    We will now demonstrate the equivalence between $f(Q)$ and $f(T)$ gravity and their counterpart with boundaries  in the flat FLRW framework. Considering the metric \eqref{flrw.metric} and adopting the  tetrad choice
    \begin{align}
        \tensor{e}{^A_\mu}={\rm diag}\bigl(1,a(t),a(t),a(t)\bigr)\,,
    \end{align}
    the only non-zero coefficients of the Weitzenb\"ock connection
    \begin{align}
        \tensor{\Gamma}{^\lambda_\mu_\nu}=\tensor{e}{_A^\lambda}\del_\nu \tensor{e}{^A_\mu}
    \end{align}
      read as 
    \begin{align}
        \tensor{\Gamma}{^i_i_0}=\frac{\dot{a}}{a}\,,
    \end{align}
    and the nonvanishing coefficients of  torsion are
    \begin{subequations}
        \begin{align}
            &\tensor{T}{^i_i_0}=-\frac{\dot{a}}{a}\,, \\
            &\tensor{T}{^i_0_i}=\frac{\dot{a}}{a}\,.
        \end{align}
    \end{subequations}
    Thus, we get the torsion scalar
    \begin{align}
        T=-6\frac{\dot{a}^2}{a^2}
    \end{align}
    and the divergence term
    \begin{align}
        \hat{B}=6\frac{\ddot{a}}{a} + 12 \frac{\dot{a}^2}{a^2}\,.
    \end{align}
    The gravitational action for the $f(T)$ gravity model in the minisuperspace $\pazocal{Q}_{f(T)}=\{a,T\}$ is
    \begin{align} \label{fT.minisuperspace.action}
        \pazocal{S}_{f(T)}=&\int\dd t a^3(t) \Biggl[f(T)-\lambda\biggl(T + 6\frac{\dot{a}^2}{a^2}\biggr)\Biggr]\,,
    \end{align}
    while in the minisuperspace $\pazocal{Q}_{f(T,\hat{B})}=\{a,T,\hat{B}\}$ we get
    \begin{align}\label{fTB.cosmological.action}
        \pazocal{S}_{f(T,\hat{B})}=&\int\dd t a^3(t) \Biggl[f(T,\hat{B})-\lambda\biggl(T + 6\frac{\dot{a}^2}{a^2}\biggr) \nn \\
        &-\rho \biggl(\hat{B} - 6\frac{\ddot{a}}{a} - 12 \frac{\dot{a}^2}{a^2}\biggr)\Biggr]\,.
    \end{align}
    These actions describe dynamics equivalent to the $f(Q)$ and $f(Q,B)$ cases, with the change of variables
    \begin{subequations}
        \begin{align}
            &T\longrightarrow-Q\,, \\
            &B\longrightarrow\hat{B}\,.
        \end{align}
    \end{subequations}
    For sake of completeness, we will show how to build the minisuperspaces for $f(T)$ and $f(T,\hat{B})$ gravity theories in Sec. \ref{subsec.fT} and \ref{subsec.fTB}. It is worth saying that this kind of equivalence between $f(Q)$ and $f(T)$ gravities is well-known at classical level, see e.g.  Ref. \cite{Wu:2024vcr}.

    \subsection{\texorpdfstring{$f(T)$ cosmology}{f(T) cosmology}} \label{subsec.fT}

        As in Sec. \ref{subsec.fQ}, we will present the cosmological model of a flat FLRW universe, along with its quantization procedure.
        
        Starting from the  gravity action in the minisuperspace $\pazocal{Q}_{f(T)}$ (see Eq. \eqref{fT.minisuperspace.action}), the variation with respect to the torsion scalar fixes  the Lagrange multiplier $\lambda$ to be
        \begin{align}\label{fT.lagrange.mult}
            \lambda=f_T(T)\,.
        \end{align}
        Plugging this result into \eqref{fT.minisuperspace.action}, we derive the Lagrangian of the $f(T)$ minisuperspace
        \begin{align} \label{initial.Lagrangian.fT}
            \pazocal{L}=a^3(f-Tf_T)-6f_Ta\dot{a}^2\,,
        \end{align}
        which permits to evaluate the Euler--Lagrange equations. Along with the energy condition $E_{\pazocal{L}}=0$, the theory is characterized by the equations of motion 
        \begin{subequations}
            \begin{align}
                &4f_Ta\ddot{a} + 2f_T\dot{a}^2 + 4f_{TT}\dot{a}\dot{T} + a^2(f-Tf_T)=0\,, \label{fT.eom.a}\\
                &T=-6\frac{\dot{a}^2}{a^2}\,,\label{fT.eom.T}\\
                &6f_T\dot{a}^2+a^2(f-Qf_Q)=0\,.\label{fT.ec}
            \end{align}
        \end{subequations}
        By substituting \eqref{fT.eom.T} into \eqref{fT.eom.a} and \eqref{fT.ec}, the  cosmological equations  finally are 
        \begin{subequations}
            \begin{align}
                &12f_T\dot{a}^2+a^2f=0\,,\\
                &12f_{TT}\dot{a}^2-a^2f_T=0\,.
            \end{align}
        \end{subequations}
        Following the same line of reasoning of previous sections, we choose the Noether vector field to be
        \begin{align}
            X=\,&\alpha(a,T)\frac{\del}{\del a} + \beta(a,T) \frac{\del}{\del T} + \dot\alpha(a,T)\frac{\del}{\del\dot a} \nn \\
            & + \dot\beta(a,T) \frac{\del}{\del\dot T}\,.
        \end{align}
        and use the identity \eqref{symmetry.definition}. By imposing that all the independent coefficients vanish, we obtain the system of PDEs 
        \begin{subequations}
            \begin{align}
                & \alpha f_T - a\beta f_{TT} - 2af_T\del_a\alpha=0\,, \label{noether.fT.eqnA}\\
                & af_T\del_T\alpha=0\,, \label{noether.fT.eqnB}\\
                & 3\alpha(f-Tf_T)+a\beta Tf_{TT}=0\,. \label{noether.fT.eqnC}
            \end{align}
        \end{subequations}
        Similarly to  $f(Q)$  case, from \eqref{noether.fT.eqnB}, we derive $\alpha=\alpha(a)$; furthermore, the ansatz
        \begin{subequations}
            \begin{align}
                &\beta=c_1\frac{T}{a}\alpha(a)+c_2T\del_a\alpha\,, \\
                &f(T)=f_0(-T)^s\,,
            \end{align}
        \end{subequations}
        with $c_1$, $c_2$ and $s$ being constants, leads to
        \begin{subequations}
            \begin{align}
                &X=\alpha_0a^{\frac{2s-3}{2s}}\del_a - \frac{3\alpha_0}{s}Ta^{-\frac{3}{2s}}\del_T\,, \\
                &j_0=-12s\,\alpha_0f_0a^{\frac{4s-3}{2s}}(-T)^{s-1}\dot{a}\,, \\
                &f(Q)=f_0(-T)^{s}\,.
            \end{align}
        \end{subequations}
        Plugging  the expression of $f(T)$ into \eqref{initial.Lagrangian.fT}, we get the Lagrangian in the $\{a,T\}$ variables
        \begin{align}\label{lagrangian.fT.Ts}
            \pazocal{L}=f_0(1-s)a^3(-T)^s+6f_0s(-T)^{s-1}a\dot{a}^2\,.
        \end{align}
        We can now choose the set of coordinates that allows us to reduce  the dynamical system. Imposing the condition \eqref{reduction.condition} to the transformation $\{a,T\}\rightarrow\{z,w\}$, we find a possible solution 
        \begin{subequations}
            \begin{align}
                & z=\frac{2s}{3\alpha_0}a^{\frac{3}{2s}}\,, \\
                & w=a^3(-T)^s\,.
            \end{align}
        \end{subequations}
        Inserting this into \eqref{lagrangian.fT.Ts}, we finally derive the Lagrangian in the reduced coordinates
        \begin{align}\label{fT.lagrangian.zw}
            \pazocal{L}=f_0w(1-s)+6f_0s\alpha_0^2w^{\frac{s-1}{s}}\dot{z}^2\,,
        \end{align}
        and the corresponding Euler--Lagrange equations
        \begin{subequations}\label{fT-zw-el-eqns}
            \begin{align}
                &sw\ddot{z}+(s-1)\dot{w}\dot{z}=0\,, \label{fT-z-el-eqn}\\
                &w=6^s\alpha_0^{2s}\dot{z}^{2s}\,. \label{fT-w-el-eqn}
            \end{align}
        \end{subequations}
        We thus immediately notice that the Lagrangian \eqref{fT.lagrangian.zw} and the equations of motion \eqref{fT-zw-el-eqns} are written in the same form of the $f(Q)$ Lagrangian \eqref{fQ.lagrangian.zw} and the $f(Q)$ equations of motion \eqref{fQ-zw-el-eqns}, thus proving the equivalence of the dynamics for $f(Q)$ and $f(T)$ gravity in the cosmological regime. This implies that the scale factor of the flat FLRW universe is \eqref{fQ.scale.factor} as well.

        The same reasoning applies to the Hamiltonian formalism and the corresponding quantum cosmological model as well, i.e. the cosmological Hamiltonian in the $f(T)$ gravity case is \eqref{fQ.canonical.Hamiltonian}, the Wheeler--DeWitt equation is \eqref{fQ.wdw}, and the Wave Function of the Universe, describing the quantized minisuperspace, is \eqref{fQ.wavefunction.universe}. This, in the semiclassical regime, leads to \eqref{fQ.scale.factor}, as expected. {A similar approach  in $f(T)$ gravity has been developed in Ref. \cite{Paliathanasis:2014iva} to seek for Schwarzschild-like solutions.

    \subsection{\texorpdfstring{$f(T,\hat{B})$ cosmology}{f(T,B) cosmology}} \label{subsec.fTB}

        Let us  now turn our attention to $f(T,\hat{B})$ and its connection with extended GR within the cosmological framework. Following the approach of Sec. \ref{subsec.fQB}, we will first use the Noether symmetries to study  $f(T,\hat{B})$ gravity, and then we will  focus on the $f(-T-\hat{B})$ restriction. 

        Starting from the action of  $\{a, T, \hat{B}\}$ minisuperspace \eqref{fTB.cosmological.action}, the variation with respect to $T$ and $\hat{B}$ provides
        \begin{subequations}\label{fTB.lagrange.mult}
            \begin{align}
                &\lambda=f_T \,\\
                &\rho=f_{\hat{B}}\,.
            \end{align}
        \end{subequations}
        Plugging \eqref{fTB.lagrange.mult} into \eqref{fTB.cosmological.action},  the point-like Lagrangian becomes
        \begin{align}\label{fTB.point.lagrangian}
            \pazocal{L}=&a^3f-a^3Tf_T-a^3\hat{B}f_{\hat{B}} \nn \\
            &- 6af_T\dot{a}^2-6a^2f_{\bt T}\dot a\dot T-6a^2f_{\bt\bt}\dot a \dot \bt\,.
        \end{align}
        The Euler--Lagrange equations are
        \begin{subequations}\label{eom.fTB}
            \begin{align}
                &f_T\dot{a}^2+2f_Ta\ddot{a} +2f_{TT}a\dot{a}\dot{T} +2f_{T\bt}a\dot{a}\dot{\bt} +2f_{\bt\bt T}a^2\dot{T}\dot{\bt} \nn \\ 
                &-f_{\bt TT}a^2\dot{T}^2 +f_{\bt T}a^2\ddot{T} +f_{\bt\bt\bt}a^2\dot{\bt}^2 +f_{\bt\bt}a^2\ddot{\bt} \nn \\
                & +\frac{a^2}{2}\bigl(f-Tf_T-\bt f_{\bt}\bigr)=0\,, \\
                &T=-6\frac{\dot{a}^2}{a^2}\,, \\
                &\bt=6\frac{\ddot{a}}{a}+12\frac{\dot{a}^2}{a^2}\,.
            \end{align}
        \end{subequations}
        The energy condition $E_{\pazocal{L}}=0$ yields
        \begin{align}
            E_\pazocal{L}=&6f_Ta\dot{a}^2 +6f_{\bt T}a^2\dot{a}\dot{T} \nn \\
            &+6f_{\bt\bt}a^2\dot{a}\dot{\bt} +a^3\bigl(f-Tf_T-\bt f_{\bt}\bigr)=0\,,
        \end{align}
        As before, we  employ the Noether symmetries to reduce the system and to select the form of the Lagrangian. By imposing \eqref{symmetry.definition} to \eqref{fTB.point.lagrangian},  the symmetry generator is
        \begin{align}
            X=&\alpha(a,T,\bt)\frac{\del}{\del a}+\beta(a,T,\bt)\frac{\del}{\del T}+\gamma(a,T,\bt)\frac{\del}{\del \bt} \nonumber\\
            &+\dot{\alpha}(a,T,\bt)\frac{\del}{\del\dot{a}}+\dot{\beta}(a,T, \bt)\frac{\del}{\del\dot{T}}+\dot{\gamma}(a,T,\bt)\frac{\del}{\del\dot{\bt}}\,.
        \end{align}
        By setting to zero all of the independent coefficients of \eqref{symmetry.definition}, we get a system of PDEs analogous to \eqref{noether.fQB.equations}, which, in this case, reads as
        \begin{subequations}\label{noether.fTB.equations}
            \begin{align}
                &f_T\alpha -af_{TT}\beta +af_{\bt T}\gamma +2af_T\del_a\alpha \nn \\
                &-a^2f_{\bt T}\del_a\beta +a^2f_{\bt\bt}\del_a\gamma =0\,, \label{noether.fTB.eqnA}\\
                &f_{\bt T}\del_T\alpha=0\,, \label{noether.fTB.eqnB}\\
                &f_{\bt\bt}\del_{\bt}\alpha=0\,, \label{noether.fTB.eqnC}\\
                &f_{\bt T}\bigl(2\alpha+a\del_a\alpha\bigr) -a\del_T\bigl(f_{\bt T}\beta -f_{\bt\bt}\gamma\bigr)=0\,, \label{noether.fTB.eqnD}\\
                &f_{\bt\bt}\bigl(2\alpha+a\del_a\alpha\bigr) -a\del_{\bt}\bigl(f_{\bt T}\beta-f_{\bt\bt}\gamma\bigr)=0\,, \label{noether.fTB.eqnE}\\
                &f_{\bt T}\del_{\bt}\alpha+f_{\bt\bt}\del_T\alpha=0\,, \label{noether.fTB.eqnF}\\
                &3\alpha\bigl(f-Tf_T-\bt f_{\bt}\bigr)\alpha +af_{TT}t\beta \nn \\
                &+af_{\bt T}\bt\beta -af_{\bt T}T\gamma-af_{\bt\bt}\bt\gamma=0\,.\label{noether.fTB.eqnG}
            \end{align}
        \end{subequations}
        
        With the ansatz
        \begin{align}\label{fTB.ansatz}
            f(T,\bt)=f(-T+\bt/k)\,,
        \end{align}
        with $k\in\R$ being a dimensionless constant, the system can be analytically solved. Having posed
        \begin{subequations}
            \begin{align}
                &\tilde{R}\defeq-T+\frac{\bt}{k}\,, \\
                &\tilde{\beta}=-\beta+\frac{\gamma}{k}\,,
            \end{align}
        \end{subequations}
        a possible solution, completely analogous to the one we found in Sec. \ref{subsec.fQB}, is then
        \begin{subequations}
            \begin{align}
                &X=\alpha_0a^{-(2+k)}\del_a -\alpha_0(3+k)\tilde{R}a^{-(3+k)}\del_{\tilde{R}}\,, \\
                &j_0=\frac{18\alpha_0f_0}{3+k}\biggl(\tilde{R}\dot{a}+\frac{a\dot{\tilde{R}}}{3+k}\biggr)a^{-(k+1)}\tilde{R}^{-\frac{3+2k}{3+k}}\,, \\
                &f(\tilde{R})=f_0\tilde{R}^{\frac{3}{3+k}}\,.
            \end{align}
        \end{subequations}
        With this reparameterization, the mapping from $f(Q,B)$ to $f(T,\bt)$ is finally complete. 
        Now, we can change variables, moving from the $\{a,\tilde{R}\}$ minisuperspace to the $\{z,w\}$ one, in which $z$ is cyclic. Posing \eqref{reduction.condition}, a possible solution is
        \begin{subequations}
            \begin{align}
                &z=\frac{a^{3+k}}{\alpha_0(3+k)}\,, \\
                &w=w_0a^\ell\tilde{R}^{\frac{\ell}{3+k}}\,,
            \end{align}
        \end{subequations}
        while the Lagrangian \eqref{fTB.point.lagrangian} can be read as
        \begin{align}\label{fTB.zw.lagrangian}
            \pazocal{L}=\frac{f_0}{(3+k)\ell w} \biggl(\frac{w}{w_0}\biggr)^{-\frac{k}{l}} \Biggl[18\alpha_0\dot{w}\dot{z} +k\ell w\biggl(\frac{w}{w_0}\biggr)^{\frac{3+k}{\ell}}\Biggr]\,.
        \end{align}
        As we anticipated, the Lagrangian \eqref{fTB.zw.lagrangian} in the reduced minisuperspace is the same as \eqref{fQB.zw.lagrangian}, thus the equations of motion, together with their solutions, are the same. In accordance with the ansatz \eqref{fTB.ansatz}, the scale factor still reads as \eqref{fQB.scalefactor}, equivalent with the $f(R)$  case  for $k=-1$.

        By performing a Legendre transform to the Lagrangian \eqref{fTB.zw.lagrangian}, we can derive the Hamiltonian 
        \begin{align}\label{fTB.zw.hamiltonian}
            \pazocal{H} = \frac{(3+k)lw}{18\alpha_0f_0}\biggl(\frac{w}{w_0}\biggr)^{\frac{k}{\ell}}\pi_z\pi_w-\frac{f_0k}{3+k}\biggl(\frac{w}{w_0}\biggr)^{\frac{3}{\ell}}\,
        \end{align}
        and, with the canonical substitutions $\pi_i\rightarrow-\ii\del_i$, we can write the Wheeler--DeWitt equation for the $\{a, \tilde{R}\}$ minisuperspace
        \begin{align}\label{wheeler.dewitt.fTB}
            \Biggl[\frac{(3+k)lw}{18\alpha_0f_0}\biggl(\frac{w}{w_0}\biggr)^{\frac{k}{\ell}}\del_z\del_w +\frac{f_0k}{3+k}\biggl(\frac{w}{w_0}\biggr)^{\frac{3}{\ell}}\Biggr]\Psi(z,w)=0\,.
        \end{align}
        With the procedure adopted in Sec. \ref{subsec.fQB}, we find that the wave function of the minisuperspace is 
        \begin{align}\label{fTB.wavefunction}
            &\Psi(z,w)= \nn \\
            &\,\,\,\Psi_0\exp\Biggl\{\ii\Biggl[j_0z + \frac{18k\alpha_0f_0^2}{(3+k)^2(3-k)j_0}\biggl(\frac{w}{w_0}\biggr)^{\frac{3-k}{\ell}}\Biggr]\Biggr\}\,,
        \end{align}
        according with the $f(Q,B)$ case. In the semiclassical limit, a comparison with \eqref{wfotu.oscill} implies that the action from the WKB approximation must read as
        \begin{align}
            \pazocal{S}=j_0z + \frac{18k\alpha_0f_0^2}{(3+k)^2(3-k)j_0}\biggl(\frac{w}{w_0}\biggr)^{\frac{3-k}{\ell}}\,.
        \end{align}
        Exploiting once again the Hartle criterion, the Hamilton--Jacobi equations for the $\{a,\tilde{R}\}$ minisuperspace must be satisfied. These are however equivalent to the equations of motion emerging from the Lagrangian \eqref{fTB.zw.lagrangian}, thus yielding the scale factor \eqref{fQB.scalefactor}, as expected. In conclusion, the equivalence between $f(R)$ gravity cosmology and $f(T,\bt)$ gravity cosmology  is restored  for $f(T,\bt)=f(-T-\bt)$.

\section{Discussion and conclusions} \label{section.conclusions}

    In this paper, we have examined the extensions of the geometric trinity of gravity  by means of the minisuperspace formalism, and we have demonstrated that the classical equivalence of $f(R)$, $f(Q-B)$ and $f(-T-\hat{B})$ theories is maintained also at quantum cosmology level. The main results are summarized in Table \ref{tab:results}. The approach relies on the Noether symmetry formalism, which allows us to constrain the dynamical structure of the system and select the gravity model. We have thus derived analytical solutions for the resulting system in both the classical and semiclassical regimes, with the semiclassical case fulfilling the Hartle criterion.

    We  first  analyzed the cosmological dynamics within the framework of $f(R)$ gravity, where the Euler--Lagrange equations yield the scale factor given by \eqref{f.R.scale.factor}. This result differs from those obtained in $f(Q)$ and $f(T)$ gravity, which, instead, predict a stiff matter-dominated universe. The equivalence among these theories can be restored by modifying the action with the total divergence terms  \eqref{fQ.divergence.term} and \eqref{fT.divergence.term}, thereby leading to  $f(Q-B)$ and $f(-T-\hat{B})$ gravity formulations.
    
    To investigate this equivalence in the quantum cosmology setting, we have applied the Noether symmetry approach to the generic $f(Q,B)$ and $f(T,\bt)$ Lagrangians. This procedure  led to  the sets of non-linear partial differential equations provided in \eqref{noether.fQB.equations} and \eqref{noether.fTB.equations}, which cannot be addressed by analytical methods in their most  general form. To overcome this difficulty, we have introduced the \emph{ansatz} $f(Q,B) = f(Q + B/k)$ and $f(-T+\bt/k)$, which allowed us to reduce the dynamical system and  demonstrate that the equivalence is recovered  for $k = -1$.

   However, it worth noticing that, in this paper,  the  equivalence among the different formulations of gravity is restored only for specific subclasses of extended theories (e.g. $f(Q-B)$, $f(-T-\bt)$, and $f(R)$). This result implies that the equivalence is  non-generic but related to particular subclasses of metric-affine gravities. This fact depends on the number of degrees of freedom of the considered models under investigation. For example, metric $f(R)$ gravity is a subclass of  $f(Q,B)$, $f(T,\bt)$,  which show a more general dynamics when the fields $Q$, $T$, $B$ and $\bt$ are considered as independent fields. A generic equivalence among different approaches has to be addressed without considering only the metric counterpart.
   
    This work brings the discussion on the geometric trinity of gravity and its extensions full circle. This line of research,  initiated in  \cite{Capozziello-DeFelice} with the identification of  Noether symmetries in $f(R)$ cosmology,  was subsequently extended in \cite{Capozziello:2022, Bajardi-2023} to include $f(T)$ and $f(Q)$ theories within the context of  minisuperspace quantum cosmology. With the addition of total divergence terms to the $f(T)$ and $f(Q)$ Lagrangians, we have now established the equivalence among $f(R)$, $f(-T-\hat{B})$, and $f(Q-B)$ at the solution level in classical and quantum cosmology, thereby aligning with the results of \cite{Capozziello-ext-trinity-2025}, where this equivalence was  proven classically at the level of  Lagrangians and field equations.

The actual conceptual novelty of the approach reported in this paper can be summarized as follows. First of all, it is worth saying that   the  equivalence among $f(R)$, $f(T)$, and $f(Q)$ theories via boundary terms are already known at  classical level. See Ref. \cite{Capozziello-ext-trinity-2025} and references therein. However,   the minisuperspace framework and the related quantum-cosmology  can be considered as a  first step towards a possible larger equivalence also at quantum level, despite  the fact that the foundation of these theories is very different. See Ref. \cite{Mancini:2025asp} for a critical discussion on this point. In other words, while the classical dynamics of these three formulations of  gravity is equivalent, differences could emerge at quantum level because the basics as Equivalence Principle, Local Lorentz Invariance and Local Position Invariance are recovered in different ways in metric, teleparallel and non-metric pictures. The fact that dynamics coincides also at minisuperspaces and  quantum cosmology can be seen as a further step to restore a full equivalence among these different theories. Clearly, as noticed above, minisuperspaces  are  "trucated" versions  and then differences could come out for the full theory.

    The approach investigated here can offer further  suggestions for future investigations. In fact, while this work focused on cosmological solutions in the flat FLRW minisuperspace,  similar analyses can be carried out for open and closed FLRW geometries or for any Bianchi model \cite{Capozziello:1996ay}. Furthermore, the system of non-linear PDEs \eqref{noether.fQB.equations} (along with its equivalent counterpart \eqref{noether.fTB.equations})  deserves to be further examined. By imposing suitable conditions, one can determine \textit{a posteriori} the form of the Lagrangian and thus the corresponding cosmological model. This perspective highlights the key role of symmetries in gravitational physics: rather than merely constraining the dynamics of a given model, Noether symmetries act as a selection criterion for the admissible form of the gravitational action itself. In this sense, symmetry principles do not just simplify the equations, but they guide the very construction of viable gravitational theories.

    \begin{widetext}
    \begin{center}
    \captionsetup{width=17.5cm} 
    \begin{minipage}{17.5cm} 
    \centering
    \renewcommand{\arraystretch}{1.8}
    \setlength{\tabcolsep}{12pt}
    \begin{tabular}{c c}
        \toprule
        Gravity Models & Solutions \\
        \midrule
        $f(R)$ & 
        $\begin{aligned}
            X_{f(R)} &= \frac{\alpha_0}{a} \partial_a - 2\alpha_0 \frac{R}{a^2} \partial_R \\
            j_{0,f(R)} &= 9\alpha_0 f_0 \biggl(R\dot{a} + \frac{1}{2} a \dot{R} \biggr) R^{-1/2} \\
            f(R) &= f_0 R^{3/2} \\
            a(t) &= a_0 \sqrt{c_0+c_1t+c_2t^2+c_3t^3+c_4t^4}
        \end{aligned}$ \\
        \midrule
        $f(Q)$ & 
        $\begin{aligned}
            X_{f(Q)} &= \frac{\alpha_0}{a^2} \partial_a - 6\alpha_0 \frac{Q}{a^3} \partial_Q \\
            j_{0,f(Q)} &= -6 \alpha_0 f_0 \frac{\dot{a}}{a \sqrt{Q}} \\
            f(Q) &= f_0 Q^{1/2} \\
            a(t) &= a_0 t^{1/3}
        \end{aligned}$ \\
        \midrule
        $f(T)$ & 
        $\begin{aligned}
            X_{f(T)} &= \frac{\alpha_0}{a^2} \partial_a - 6\alpha_0 \frac{T}{a^3} \partial_T \\
            j_{0,f(T)} &= -6 \alpha_0 f_0 \frac{\dot{a}}{a \sqrt{-T}} \\
            f(T) &= f_0 (-T)^{1/2} \\
            a(t) &= a_0 t^{1/3}
        \end{aligned}$ \\
        \midrule
        $f(Q+B/k)$ & 
        $\begin{aligned}
            X_{f(Q+B/k)} &= \alpha_0 a^{-(2+k)}\partial_a -\alpha_0(3+k)(Q+B/k)a^{-(3+k)}\partial_{Q+B/k} \\
            j_{0,f((Q+B/k))} &= \frac{18\alpha_0 f_0}{3+k} 
            \biggl((Q+B/k) \dot{a} + \frac{a (\dot{Q}+\dot{B}/k)}{3+k} \biggr) 
            a^{-(k+1)}(Q+B/k)^{-\frac{3+2k}{3+k}} \\
            f(Q+B/k) &= f_0 (Q+B/k)^{\frac{3}{3+k}} \\
            a(t) &=\biggl[a_0+a'_0t -\frac{k^3(k+3)}{18(k-3)}(t-t_0)^{\frac{k-3}{k}}\biggl]^{\frac{1}{k+3}}
        \end{aligned}$ \\
        \midrule
        $f(-T+\bt/k)$ & 
        $\begin{aligned}
            X_{f(-T+\bt/k)} &= \alpha_0 a^{-(2+k)}\partial_a -\alpha_0(3+k)(-T+\bt/k)a^{-(3+k)}\partial_{-T+\bt/k} \\
            j_{0,f(-T+\bt/k)} &= \frac{18\alpha_0 f_0}{3+k} 
            \biggl((-T+\bt/k) \dot{a} + \frac{a (-\dot{T}+\dot{\bt}/k)}{3+k} \biggr) 
            a^{-(k+1)}(-T+\bt/k)^{-\frac{3+2k}{3+k}} \\
            f(-T+\bt/k) &= f_0 (-T+\bt/k)^{\frac{3}{3+k}} \\
            a(t) &=\biggl[a_0+a'_0t -\frac{k^3(k+3)}{18(k-3)}(t-t_0)^{\frac{k-3}{k}}\biggl]^{\frac{1}{k+3}}
        \end{aligned}$ \\
        \bottomrule
    \end{tabular}
    \vspace{0.5em}
    \captionof{table}{Comparison among different  gravity models. For each theory,  the Noether vector field $X$,  the conserved charge $j_0$, the Lagrangian density,  selected by the Noether symmetry,  and the ensuing scale factor $a(t)$ are reported. $f(R)$ cosmology is compatible for large $t$ with a dark energy scenario, while the equivalent $f(Q)$ and $f(T)$ cosmologies lead to a stiff matter-dominated universe. The correspondence with $f(R)$ cosmology is restored for $f(Q+B/k)$  and $f(-T+\bt/k)$ gravity in the case $k=-1$.}
    \label{tab:results}
    \end{minipage}
    \end{center}
    \end{widetext}

\section*{Acknowledgements}
  The authors   acknowledge the support of  Istituto Nazionale di Fisica Nucleare (INFN) Sezioni  di Napoli e di Frascati, {\it Iniziative Specifiche} QGSKY and MOONLIGHT2. SC thanks the Gruppo Nazionale di Fisica Matematica (GNFM)  of Istituto Nazionale di Alta Matematica (INDAM) for the support.   
  This paper is based upon work from COST Action CA21136 {\it Addressing observational tensions in cosmology with systematics  and fundamental physics} (CosmoVerse) supported by COST (European Cooperation in Science and Technology). 

\appendix

\bibliography{references}

\end{document}